\documentclass[letterpaper,12pt]{article}
\usepackage{setspace}
\usepackage[left=1.25in, top=1.25in, right=1.25in, bottom=1.25in]{geometry}
\usepackage{adjustbox}
\usepackage[charter]{mathdesign}
\usepackage{XCharter}
\usepackage[utf8]{inputenc}
\usepackage{amsmath}
\usepackage{amsthm}
\usepackage{appendix}
\usepackage{xr-hyper}
\usepackage{hyperref}
\usepackage{booktabs,threeparttable,tabularx,makecell}
\usepackage{subcaption}
\usepackage{csquotes}
\usepackage{enumerate}
\usepackage{float}
\usepackage{graphicx}
\usepackage{placeins}
\usepackage[table,xcdraw]{xcolor}
\usepackage{bm}
\usepackage{dsfont} %
\usepackage{ulem} %
\usepackage{xcolor}
\usepackage[backend=biber,
	authordate,
	url=false,
	doi=false,
	isbn=false,
	eprint=false]{biblatex-chicago}
 \usepackage{pgfplots}
 \pgfplotsset{compat=newest}
 \usepackage{tikz}
 \usetikzlibrary{calc, patterns}


\newcounter{parentnumber}

\theoremstyle{plain}%

\newtheorem{theorem}{Theorem}
\newtheorem{regularity}{Regularity}
\newtheorem{proposition}{Proposition}
\newtheorem*{proposition*}{Proposition}
\newtheorem{lemma}{Lemma}
\newtheorem*{lemma*}{Lemma}
\newtheorem{corollary}{Corollary}
\newtheorem*{corollary*}{Corollary}
\newtheorem{condition}{Condition}

\theoremstyle{definition}
\newtheorem{definition}{Definition}
\newtheorem{example}{Example}
\newtheorem{axiom}{Axiom}
\newtheorem*{RC}{Regularity Condition}

\DeclareMathOperator{\supp}{supp}

\title{Non-Allais Paradox and Context-Dependent Risk Attitudes \footnotetext{The authors thank Soo Hong Chew, Shachar Kariv, Paulo Natenzon, John Rehbeck, Brian Rogers, Jason Somerville, Norio Takeoka, Gerelt Tserenjigmid, Julia Witt, Daeyoung Jeong, Euncheol Shin, and audiences at BEEMA (NY Federal Reserve Bank), ESA (Santa Barbara \& Seoul), MEA (Cleveland), NASMES (Nashville), WEAI (Seattle), and ESI Seminar (Chapman University) for helpful discussions and comments. We give special thanks to Pete Caradonna for suggesting the current version of our model that generalizes our initial version in a previous draft. We are grateful for the administrative and financial support from the Smith Institute for Political Economy and Philosophy of Chapman University.}}
\author{
  Edward Honda\footnote{Department of Economics, University of Manitoba. Email: Edward.Honda@umanitoba.ca}
  \and
  Keh-Kuan Sun\footnote{Department of Economics, Fairfield University.  Email: ksun@fairfield.edu}
}

\date{March 10 2026}

\begin{document}
   	\maketitle\small
   	\begin{abstract}
   	
 	 We provide and axiomatize a representation for preferences over lotteries that generalizes the expected utility model. Since the representation uses different utility functions to evaluate different lotteries, the preferences can be interpreted as coming from individuals that have context-dependent attitudes toward risks. The model enables generating various violations of the independence axiom that are not compatible with some of the most prominent models of non-expected utility. Depending on the specification chosen, the model can range from being very flexible with many different utility functions to being parsimonious with few or just one utility function.
 	
     \vspace{2em}

    
   	\end{abstract}

    \vfill
    \pagebreak

\section{Introduction}
\label{sec:introduction}
Expected utility (EU) has been and still is the main workhorse in economic modeling in a wide range of applications. The number of alternative theories proposed and empirical evidence against the theory, however, have only accumulated over decades. The famous examples provided by \textcite{Allais1953} are now known to many as the Allais Paradox and motivated earlier non-EU models. More recently, newer studies continue discovering systematic departures from the EU predictions in experiment designs extended beyond the traditional Allais patterns. Our paper adds another non-Allais regularity, the worst-outcome-shift (WS) pattern, where preference reversals are triggered by shifting the worst outcome while holding probabilities fixed.

The purpose of our work is to develop an axiomatic model of decision making under risk that (i) nests EU as a special case; (ii) is consistent with the Allais and various non-Allais behavioral patterns; and (iii) yields testable, falsifiable restrictions beyond reduced‑form flexibility. We do so by relaxing the independence axiom to account for the \emph{context} of a lottery, captured by the probability of extremely low outcomes. In our model, the utility function the decision maker (DM) uses to evaluate a lottery depends on the context of this lottery. Since different utility functions can be associated with different risk attitudes, the variation in the utility functions allows changes in the risk attitude of the DM to be a part of the representation. This enables us to generate various types of realistic behavior that seem to arise from changes in risk attitudes of the DM but are not consistent with some of the prominent models that generalize EU theory.

We illustrate this point with two numerical examples. First, consider the following WS problem. Suppose a DM who faces a risky prospect in which there is a high probability of ending up with a low prize exhibits the following risk-loving preference, $$(\$0, 0.9; \$100, 0.05; \$200, 0.05) \succ  (\$0, 0.9; \$150, 0.1).$$ When we swap the \$0 in both lotteries with a prize value of $\$90$ to ensure a moderate level of wealth though, the DM becomes risk averse and exhibits the preference $$(\$90, 0.9; \$100, 0.05;\$200, 0.05) \prec (\$90,0.9; \$150, 0.1).$$ This example can easily be tested in an experimental setting. We first provide a pair of two lotteries which have a common worst outcome. We then shift this outcome by replacing the worst outcome with a higher value and observe that subjects exhibit a preference reversal. Since we simply shifted the worst outcomes while the probabilities or ranks associated with the outcomes remain unchanged, this example is clearly not consistent with rank-dependent probability weighting models such as cumulative prospect theory (CPT) \autocite{Kahneman1979,Quiggin1982,Tversky1992}.\footnote{Our example is not consistent with CPT if we let \$0 be the reference point for the model, but we can construct a similar example for any other reference point as well.} Also, the example is not consistent with disappointment aversion, generalized disappointment aversion, or cautious expected utility model \autocite{Gul1991,routledge2010,Cerreia2015}.\footnote{Appendix \ref{app:ws-non-eu} derives benchmark DA, GDA and CEU predictions for the WS family and shows that 1) under any parameterization, the WS value difference $\Delta(w)$ can change sign at most once; 2) under standard risk-averse specifications, no reversal occurs over the experimental grid.} 

Consider another example with a mixture of lotteries. When offered the lottery $p=(\$100,0.5;\$20,0.5)$, the DM considers the 0.5 probability on the lower prize of \$20 to be disturbingly high so prefers the less risky lottery $q=(\$60,1)$. However, when the riskier lottery is altered to $r=(\$100,0.2;\$60,0.6;\$20,0.2)$, the lowered probability of getting the disappointing \$20 causes the DM to consider it negligible. This encourages the DM to become risk-loving and want to take a chance on achieving the highest prize of \$100. As a result, the DM prefers this lottery over $(\$60,1)$. This is a violation of the betweenness axiom because the lottery $r$ is actually a mixture of $p$ and $q$. Therefore, it is not consistent with any model that has betweenness as a necessary condition. Moving from $p$ to the mixture $r$ lowers the probability mass on the worst outcome, which can move evaluation into a different context and thereby flip local risk attitude, even though $r$ is a mixture of $p$ and $q$.
Table \ref{tab:nonEU_comparison} summarizes the leading non-EU models and their predictions with respect to these examples.

Both of these examples of violations of independence are similar to the Allais Paradox in the sense that they seem to be results of changes in the risk attitude that stems from changes in the contexts of the lottery. Yet, the fact that the modifications being made to the lotteries are not identical to the ones being made in the Allais Paradox suddenly make the preferences inconsistent with some of the popular models of non-EU. One of the key features of our model is that it is consistent with various types of non-Allais violations of independence that arise from changing risk attitudes and not just the behavior in the Allais Paradox. 

\begin{table}[htbp]
\centering
\footnotesize
\begin{adjustbox}{rotate=90, max width=\textheight}
\begin{threeparttable}
    \setlength{\tabcolsep}{6pt}
    \renewcommand{\arraystretch}{1.5}
    
    \newcolumntype{L}{>{\raggedright\arraybackslash}X}
    \newcolumntype{S}{>{\centering\arraybackslash}p{5em}}
    \begin{tabularx}{\textheight}{@{} L r L L S S S @{}}
    \toprule
    \thead{Model} & \thead{Year} & 
    \thead{Representation} & \thead{Key Innovation} & 
    \thead{Non-Allais\\ patterns} &
    \thead{Multiple switch} \\
    \midrule
    Expected Utility & 1947 & $\sum p(x)\,u(x)$
    & Baseline & No & No \\
    Weighted Utility & 1977 & $\sum w\!\big(p(x)\big)\,u(x)$
    & Weight probabilities & No & No \\
    Prospect Theory & 1979 & $\sum w\!\big(p(x)\big)\,v(x)$
    & Weight{+}value function & No & No \\
    Rank-Dependent Utility & 1982 & $\sum \pi_i\,u(x_i)$
    & Rank-dependent weights & No & No \\
    Disappointment Aversion & 1991 & $V(p)$ implicitly defined through the CE of $p$
    & Endogenous reference; weak independence & Some\tnote{*} & No \\
    Generalized Disappointment Aversion & 2010 &
$u(CE(p))=\mathbb{E}[u(x)]-\theta\,\mathbb{E}\![(u(\delta\,CE(p)) -u(x))\mathbf{1}\{x\le \delta\,CE(p)\}]$ &
Lottery-dependent internal threshold $\delta\,CE(p)$; state-dependent DA in applications & Some\tnote{*} & No \\
    Cautious EU & 2015 & $\displaystyle \inf_v \left\{ \sum p(x)\,v(x) \right\}$
    & Multiple utilities; negative certainty independence & Some\tnote{*} & No \\
    Upside potential & 2025 & $\sum p_i\,u(x_i) + f\!\Big(\sum_{j\in W} p_j\Big)\,\frac{\sum_{k\in W} p_k\,\phi(x_k)}{\sum_{j\in W} p_j}$
    & EU + upside-potential bonus term\tnote{\dag\dag} & Some\tnote{*} & No \\
    \midrule
    Expected Contextual Utility & 2025 & $\sum p(x)\,u_{\pi}(x)$
    & Context-dependent utility; contextual substitutability & Yes & Some\tnote{\dag} \\
    \bottomrule
    \end{tabularx}
    
    \caption{Comparison of Non-EU Models}
    \label{tab:nonEU_comparison}
    
    \begin{tablenotes}[para,flushleft]
    \item \textit{Notes:} Non-Allais patterns refer to systematic violations of EU/independence beyond the classical Allais CR/CC comparisons, including the WS operation, the mixture/betweenness pattern highlighted in the text, and connected CR--CC--MX patterns.
    
    \item[*] Entries marked with * depend on parameter restrictions; see Appendix~\ref{app:crccmx-ecu} for CR--CC--MX sign predictions under ECU and Appendix~\ref{app:ws-non-eu} for WS benchmarks (DA, GDA, CEU, UP). 
    
    \item[\dag] We discuss how ECU can accommodate some types of regular multiple switching behavior in Section~\ref{sec:remark}.

    \item[\dag\dag] $W$ denotes the set of winning outcomes; $\kappa$ is a strictly increasing upside-potential function.

    \end{tablenotes}
\end{threeparttable}
\end{adjustbox}
\end{table}

We refer to our representation as an Expected Contextual Utility (ECU) representation. The model reduces to the EU when the DM has only one utility function to evaluate the lotteries. The representation is also unique up to affine transformation and, as in the original EU, satisfies preference for FOSD under mild conditions. While the model can be very flexible as it allows an infinite number of different utility functions to be used, it can be parsimonious if we use only a small number of distinct functions. Such restriction also makes it easier to verify whether the testable implications of the model are satisfied or not. For these reasons, we provide and carefully analyze a special case in which it is as if the DM uses only two distinct utility functions. In this special case that we call the Binary ECU, the DM uses the same utility function for all lotteries with contexts where the probability of disappointment is below a particular threshold, and uses another one for all other lotteries that have levels of disappointment greater than the threshold. It turns out that even this simple model is flexible enough to accommodate behaviors of our interest. Since we use only two functions, the complexity required is comparable to CPT, which also uses two functions (a utility function and a weighting function). More generally, in Section 3 we show that the full model can be parameterized by expressing infinitely many utility functions through a single parameter, achieving flexibility comparable to other non-EU models without additional complexity.

Aside from the theoretical component of our work, the methods used in and the results of the experiment may be of independent interest. Much of the literature on experiments involving violations of the independence axiom focuses on the Allais Paradox, whereas we find violations that are distinct from those associated with the Allais Paradox. These violations are inconsistent with many leading non-EU models.

The rest of our work proceeds as follows. Section 2 places our work in the related literature. In Section 3, we formally define our ECU representation and introduce noteworthy special cases. We discuss the range of behaviors that our model generates in Section 4. Section 5 provides the axiomatic characterization and the uniqueness result for our model. In Section 6, we describe the experimental design and report the main results of the experiment. We conclude with Section 7. All proofs are relegated to the Appendix.

\section{Related Literature and Experimental Evidence}
Since \textcite{Allais1953}, a large literature has proposed alternatives to EU to account for the common ratio (CR) and common consequence (CC) problems. A leading example is Prospect Theory proposed by \textcite{Kahneman1979} that addresses probability distortion and reference dependence through probability weighting and a gain-loss value function. CPT \autocite{Tversky1992} incorporates rank-dependent probability weighting \autocite{Quiggin1982} to restore first-order stochastic dominance and remains a leading behavioral model.

Despite its intuitive appeal, however, the empirical evidence for the CPT is mixed. Most notably, \textcite{Bernheim2020} find that relative decision weights barely respond to rank changes and conclude that probability weighting is non-linear but essentially rank-independent. The result undermines the core concept of CPT and calls for a rank-independent non-EU model. A complementary strand emphasizes a different invariance test: manipulating the representation of a lottery by splitting or coalescing branches that yield the same
outcome. \textcite{birnbaum2006,birnbaum2007,birnbaum2008} shows systematic violations of coalescing (event-splitting effects), which also contradict the rank-dependent machinery of CPT/RDU but through changes in branch structure rather than outcome levels. Our ECU framework is rank-independent by design: probabilities remain linear and non-EU behavior arises through context-dependent changes in utility. Our simple experiment of the WS problem provides complementary evidence to their finding by documenting preference reversal induced by rank-independent shifts in the prizes of lotteries. We also note that our WS design is orthogonal to the branch splitting invariance in that it holds the branch structure (and hence ranks and probabilities) fixed while shifting the common worst outcome.

A separate axiomatic line of models that replace independence with a weaker axiom of betweenness, such as \textcite{Dekel1986, chew1989, Gul1991, routledge2010}, form the most prominent classes for early axiomatic models of non-EU. In particular, the disappointment-aversion (DA) discussed in \textcite{Gul1991} is closest to ours among these three models.\footnote{Along with Gul's DA, the generalized disappointment aversion model of \textcite{routledge2010} replaces the certainty-equivalent cutoff with a scaled cutoff $\delta\,CE(p)$; for our purposes the distinction is mainly notational, and the WS benchmark logic applies with only minor modifications (see Appendix~\ref{app:ws-non-eu}).} As stated in the introduction, the main objectives of our approach are heavily influenced by that of Gul: both DA and ECU seek to axiomatize a representation that generalizes the EU while providing more flexibility. However, an important distinction is that ECU inverts the mechanics of Gul by pairing a fixed disappointment threshold $d$ with multiple utility functions, instead of a lottery-dependent threshold with a single fixed utility function. Naturally, this modeling choice requires distinct axiomatic structure and yields divergent behavioral implications.

Inasmuch as they do not distort the weights on prizes and incorporate multiple utility functions, the cautious expected utility of \textcite{Cerreia2015} and the models of \textcite{Maccheroni2002} and \textcite{Cerreia2009} are similar to our model. There are, however, differences in terms of both the interpretation and the behaviors generated by the models. These models were developed in response to the certainty effects, one of the most robust drivers of independence violations. In particular, \textcite{Cerreia2015} build on the Negative Certainty Independence axiom (NCI, \textcite{Dillenberger2010}), which is a weaker condition than the independence axiom and necessary for cautious expected utility. They posit as if the DM chooses among multiple utility functions and computes the EU of a lottery for all of those functions when evaluating this lottery. When assigning a value to any lottery after considering all of the utility functions, the DM chooses the evaluation that is the most cautious or pessimistic. This property of evaluating multiple utility functions explains more frequent independence violation near certainty. By contrast, for the ECU representation that satisfies preferences for first-order stochastic dominance, the direction of pessimism can vary with the likelihood of disappointment.

Subsequent empirical findings, however, cast doubt on these axiomatized generalizations of EU. \textcite{camerer1994} and \textcite{blavatskyy2006} summarize early experiments that show frequent violations of the betweenness axiom. \textcite{jain2024a} find a substantial share of independence violations in their study result from the reverse certainty effect, which is a violation of NCI. Complementing this, \textcite{incekara-hafalir2021} document a robust zero effect, whereby individuals display strong aversion to receiving nothing.  This pattern aligns with the motivation for our ECU framework: the behavioral switch is triggered when the probability mass on ``disappointing'' outcomes increases. Accordingly, we note that the ECU can violate both the betweenness and the negative certainty independence axioms and we discuss them in detail in Appendix  \ref{app:ws-non-eu}.In a complementary direction, \textcite{dembo2025} use a three-dimensional budget-set experiment to decompose EU violations nonparametrically and find that, for the vast majority of subjects, failures of ordering and FOSD monotonicity dominate failures of independence. We return to this finding in Section~\ref{sec:remark}.

More recently, \textcite{mcgranaghan2025} connect CR-CC-MX problems and report patterns that are hard to reconcile with betweenness-based models. The modal behavioral pattern they observe exhibits CR preference, reverse CC in a certain parameter region, and a robust attraction to probabilistic mixtures over wide parameter ranges. In response to the empirical characteristics they propose a model that supplements the standard EU functional with a ``upside-potential'' term, which captures the total probability of winning and the conditional expected value given a win. Both the ECU model and the upside-potential model are rank-independent and directly respond to the nonparametric evidence against rank dependence in \textcite{Bernheim2020}. Also, conceptually, both models partition lottery spaces into `win' versus `bad/disappointing.' Thus the two models can yield similar qualitative comparative statics and be complementary with each other while differ in primitives and restrictions. To illustrate the compatibility of the two models, we discuss how the binary ECU can reproduce the modal behavioral patterns of \textcite{mcgranaghan2025} in Appendix \ref{app:crccmx-ecu}.

The main intuition behind the ECU framework for explaining changing risk attitude through variation across contexts has existed in literature in various forms. For instance, \textcite{friedman_savage_1948} proposed curvature changes over wealth, maintaining the independence axiom. Later, lottery-dependent utility models such as \textcite{Becker1987} and \textcite{Schmidt2001} made the idea more explicit by allowing utility to vary for each lottery being evaluated. However, these models have seen limited use in applications perhaps because they rely on a priori knowledge about how the DM separates the set of lotteries, as well as empirical challenges to certain regulatory conditions they impose. The ECU framework formalizes this insight of context dependence in a modern, axiomatized way and better explains recent empirical patterns in a testable structure.

\section{Expected Contextual Utility Representation}
\subsection{Defining the Representation}
We start this section by introducing the setup for the model. We let $X=[w,b]\subset \mathbb{R}$, where $w<b$, represent a set of prizes. Endow $X$ and $\mathbb{R}$ with a metric topology where the metric is $dist(x,y)=|x-y|$. The set of Borel probability measures with finite support on $X$, which we also call simple lotteries, is denoted $\mathcal{L}$. For any $x\in X$, $\delta_x$ is the Dirac measure at $x$. We let $\succsim$ be a binary preference relation over $\mathcal{L}$. 

We define the threshold for what the DM considers as disappointing prizes as in \textcite{Honda2022}. The value $d\in [w,b]$ is the threshold for the DM's disappointment with a prize, so prizes in $[w,d]$ are disappointing. The utility functions that the DM uses to evaluate a lottery depends on the probability associated with disappointing prizes, or prizes that are less than or equal to $d$. 

Thus, for each $\pi\in (0,1)$, we have a function $u_\pi:X\rightarrow \mathbb{R}$ that can be interpreted as a utility function over prizes. If a lottery $p\in \mathcal{L}$ assigned probability $\pi\in(0,1)$ to prizes in $[w,d]$ or $p([w,d])=\pi$, the DM uses the utility function $u_{\pi}$. This allows for changes in risk attitudes based on a lottery's context. We can thus generate behaviors like those mentioned in the introduction with risk-lovingness for high probabilities of disappointment and risk-aversion for low probabilities. 

Note that we have not yet defined utility functions for $u_0$ and $u_1$  because their domains differ slightly. Based on our description thus far, function $u_0$ is used to evaluate lotteries that place probability 0 on disappointing prizes, meaning that disappointing prizes are not in the support. That being the case, utility values for disappointing prizes do not need to be defined for $u_0$, and we define function $u_0:X\setminus [w,d]\rightarrow \mathbb{R}$. 

In similar fashion, $u_1$ is used for lotteries that offer only disappointing prizes in the support. For this reason, the utility values $u_1(x)$ do not need to be defined for $x\in(d,b]$. As such, we define function $u_1:[w,d]\rightarrow \mathbb{R}$. We could also adopt an alternate approach where we extend the domains of $u_0$ and $u_1$ to $[w,d]$. In this case the values of $u_0(x)$ and $u_1(y)$ are irrelevant for $x\in [w,d]$ and $y\in X\setminus [w,d]$ and would lead to the lack of any type of uniqueness of the representation because these values can be defined arbitrarily as they have no relevance when evaluating values of lotteries. We avoid using this approach as it unnecessarily complicates much of the following discussion.\footnote{We do, however, adopt this approach occasionally in Section 3 for convenience when we introduce special cases.}

We impose several mild regularity conditions on the utility functions. For any $\pi\in[0,1]$ such that $u_{\pi}(w), u_{\pi}(b)$ are defined\footnote{\(u_0 (x)\) is not defined for $x\in [w,d]$, and \(u_1 (x)\) is not defined for $x\in X\setminus [w,d]$. }, we impose $u_\pi(w)< u_\pi(b) $. This specification simply implies that the utility the worst prize provides is lower than the utility the best prize provides. Next, for any $x\in\{w,b\}$ and $\pi,\mu\in[0,1]$ such that $u_{\pi}(x), u_{\mu}(x)$ are defined, $u_\pi(x)=u_\mu(x) $. In simple terms, all utility functions that contain the worst prize in their domains will evaluate it identically, and, similarly, all utility functions that include the best prize in their domains evaluate it identically. For example, $u_0$ will not include $w$ in the domain but all other utility functions must be defined on $w$, which means we must have $u_\pi(w)=u_\mu(w)$ for all $\pi,\mu\in (0,1]$. 

We sometimes refer to the common value for the utility of the worst prize as simply $u(w)$ and omit the subscript that indexes the utility functions. We do the same for the best prize and sometimes write $u(b)$. We then impose the condition that $u_{\pi}(x)\in [u(w),u(b)]$ for all $x\in(w,b)$ and $\pi\in [0,1]$ such that $u_\pi(x)$ is defined.

Finally, we impose the condition that, if $d\neq b$, then for any $x\in (w,b)$ there exist $\pi,\mu\in (0,1)$ such that $u_\pi(x)\neq u_\mu(x)$. To understand what this condition implies, note that we have a standard EU maximizer when the DM uses the same utility function to evaluate all lotteries. This would happen if $d=b$, as all lotteries would assign  probability 1 to disappointing prizes. For other cases, the DM uses multiple utility functions, and the condition ensures that there is sufficient variation in the way that a prize is evaluated.  

We say that a set of functions ${\mathcal{U}}=\{u_\pi| \pi\in [0,1]\}$, consisting of real valued functions, that satisfies all these properties is \textbf{contextual}.
 	
    \begin{definition}
        A preference relation $\succsim$ has an expected contextual utility (ECU) representation iff there exists $d\in [w,b]$ and a set of utility functions $\mathcal{U}$ that is contextual, such that, for any $p,q\in \mathcal{L}$, $p\succsim q$ iff $V(p)\geq V(q)$, where for any $r\in \mathcal{L}$, if $r([w,d])=\pi$,
        \[
            V(r)=\int u_\pi \; dr.
        \]
    \end{definition}

 \subsection{Discussion on the Disappointment Parameter}
As noted previously, a key characteristic that differentiates ECU from many existing models is that the threshold $d$ is universal for all lotteries. What the DM considers to be insignificant prizes are not affected by the specific lottery under consideration. This suggests that we can interpret the parameter $d$ as a fixed threshold of a DM who has a clear sense of what is a small or large amount of money. Thus, one way to interpret the variation in $d$ across different DMs is to treat the variation as coming from differences in their socioeconomic status. For instance, a DM with a high income level may have a high threshold since they will dismiss a lot of monetary values as being insignificant compared to the amount that they are used to receiving. For such individuals $\$5$ should certainly be considered a small prize in a lottery that may give a prize of $\$5000$ with positive probability. But even if another lottery consists of prizes only between $\$0$ and $\$5$ so that $\$5$  is now the best prize, this should not suddenly make the same DM with a fixed socioeconomic status view $\$5$ as a high payoff. This feature is desirable for the type of behavior we have in mind in  which the DM is risk-loving  when disappointments are very likely but risk-averse when disappointments are less likely. 

To understand this point, consider lotteries that give the low prize of $\$5$ with a high probability of 0.99 but gives prizes in the range $[4500,5000]$ with probability 0.01. There is a high probability of disappointment, so the DM should be risk-loving for these lotteries. Now, consider  lotteries that give $\$5$ with a high probability of 0.99 again, but places the rest of the probability on prizes less than $\$5$. If we make the disappointment threshold dependent on the lottery, $\$5$ would now be considered a non-disappointing prize because it is the best prize in all of these lotteries. Since there is a high probability of this maximal prize of $\$5$, the probability of disappointment for such lotteries would be considered to be low and the DM may end up using utility functions associated to risk-aversion. However, in reality, we would expect the DM to be risk-loving for such insignificant lotteries involving low stakes because $\$5$ is just $\$5$ regardless of the values of other relevant prizes.

\subsection{Some Special Cases}
With the formal definition in hand, we proceed to introducing some special cases of the ECU to show how it can have a wide range of specifications. We start with the case that ensures preference for first order stochastic dominance, a normatively appealing property that, as in the original EU, requires mild regularity conditions.

\subsubsection{Preference for Stochastic Dominance}
For the sake of generality, we have not imposed many restrictions on the utility functions up to this point, but we now introduce restrictions on the ECU that ensure preferences for first-order stochastic dominance. We use the standard definition for first-order stochastic dominance.

    \begin{definition}
        Lottery $p\in \mathcal{L}$ \textbf{first-order stochastically dominates} lottery $q\in \mathcal{L}$ iff for any $x\in X$, $p([w,x])\leq q([w,x]).$
    \end{definition}

To ensure that first-order stochastically dominant lotteries are weakly preferred over those they dominate, we impose the following conditions on the set of utilities in addition to requiring them to be contextual.
    \begin{condition} 
        \label{condition:non-decreasing-u}
        $u_\pi$ is non-decreasing for all $\pi\in[0,1]$. 
    \end{condition}
    \begin{condition}
        \label{condition:monotone-pessimism}
        $u_{\pi}(x)\geq u_{\mu}(x)$ for any $x\in X$ and $0\leq \pi\leq\mu\leq 1$ such that $u_{\pi}(x), u_{\mu}(x)$ are defined.
    \end{condition} 
The second condition says that the DM becomes more pessimistic and assign lower values to prizes as disappointments become more likely. This pessimism ensures that the represented preferences satisfy preferences for first-order stochastic dominance. 

    \begin{proposition}
        If $\succsim$ has an ECU representation with a set of contextual functions that satisfies Conditions \ref{condition:non-decreasing-u} and \ref{condition:monotone-pessimism}, then $p\succsim q$ for any $p,q\in \mathcal{L}$ such that $p$ first-order stochastically dominates $q$.
    \end{proposition}

    \begin{proof}
        See Appendix \ref{appendix:proofs}.
    \end{proof}

\subsubsection{Binary Utilities}
In principle, the ECU allows a distinct utility function for each value of the probability that disappointment occurs. This means that a DM may use infinitely many utility functions. It could also be the case that the same utility function is used for many values of disappointment probability and the DM uses only a small and finite number of utility functions.

As suggested in the Introduction, we provide a very simple special case in which the DM only has two distinct utility functions. It will be as if the DM uses one utility function for all lotteries that are considered to have \enquote{low} probabilities of disappointment and another for those that have \enquote{high} probability of disappointment. We provide this special case because this simple example turns out to be enough to generate a lot of the behavior that we want to capture. More of this is discussed in Section \ref{sec:resulting_behavior}. The special case satisfies Conditions \ref{condition:non-decreasing-u} and \ref{condition:monotone-pessimism} so that it is consistent with preference for first-order stochastic dominance.

We introduce a value $\tau \in [0,1] $ that serves as a threshold level of probability for what the DM considers to be a low level of disappointment. If a lottery assigns a probability that is less than or equal to $\tau$ to prizes whose values are less than or equal to $d$, the DM evaluates the lottery using one strictly increasing utility function $u:X\rightarrow \mathbb{R}$. When the probability that disappointing prizes are awarded exceeds $\tau$, this becomes a concern for the DM and causes a change in risk attitude. Formally, there is a strictly increasing function $v:X \rightarrow \mathbb{R}$, and the DM evaluates a lottery using this utility function when $p([w,d])>\tau$. Condition \ref{condition:non-decreasing-u} is satisfied because the two functions are strictly increasing. 

To the extent that we have an ECU, we need $v(w)=u(w)$ and  $v(b)=u(b)$. We also impose $v(x)<u(x)$ for $x\in(w,b)$ to ensure that the two utility functions constitute a contextual set and that Condition \ref{condition:monotone-pessimism} is satisfied. Note that $u$ plays the role of $u_\alpha$ for all $\alpha\in [0,\tau]$.  We call this a \textbf{binary ECU}.

\subsubsection{Disappointment Probability as Utility Parameters} 
 Here, we give another special case of the ECU to show that we can have an ECU with an infinite number of utility functions that can be expressed in an extremely simple manner. It uses just the parameter value $\pi$ of the probability placed on disappointing prizes. 

For any $\pi\in [0,1]$ and $x\in X$, we let $$u_\pi(x)=\left(\frac{x-w}{b-w}\right)^{0.5 + \pi}.$$ To see that this set of utility functions is contextual, note that $u_\pi(w)=0$ and $u_\pi(b)=1$ for all $\pi \in[0,1]$. It is also the case that $u_\pi$ is strictly increasing for all $\pi \in [0,1]$ so that Condition \ref{condition:non-decreasing-u} is satisfied along with the condition that $u_\pi(x)\in[ u(w), u(b)]$ for all $x$ and $\pi$ such that $u_\pi(x)$ is defined. Lastly, for any $0\leq \pi<\mu\leq 1$, and $x\in (w,b)$ we know that $u_\pi(x)>u_\mu(x)$ so that Condition \ref{condition:monotone-pessimism} is satisfied, ensuring that, for any $x\in (w,b)$, there exists $\pi,\mu\in [0,1]$ such that $u_\pi(x)\neq u_\mu(x)$.

     \begin{figure}[ht]
        \centering
        \includegraphics[width=0.6\textwidth]{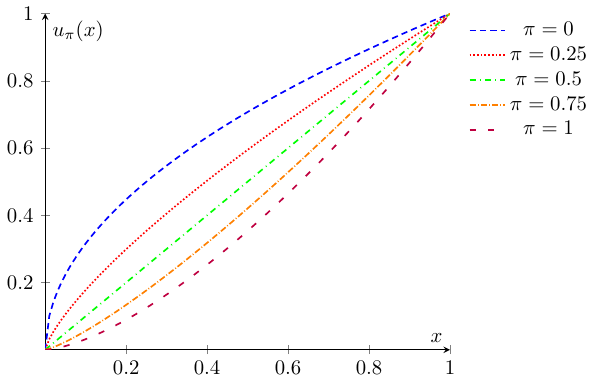}
        \caption{\(u_\pi(x)=\left(\frac{x-w}{b-w}\right)^{0.5 +\pi }\) with \(w=0, b=1\) }
        \label{fig:smooth_parametric_example}
    \end{figure}

Note that we constructed the functional form so that the DM's risk attitude changes from risk-averse to risk-loving after a certain threshold as the probability placed on disappointing prizes increases. In Figure \ref{fig:smooth_parametric_example}, for example, the utility function is concave when \( \pi < 0.5 \), linear when \( \pi = 0.5 \), and convex when \( \pi > 0.5 \).

\section{Resulting Behavior}
\label{sec:resulting_behavior}
Our model allows us to accommodate many types of realistic preferences that seemingly stem from  changing risk attitudes but may not be consistent with some popular models. We give some examples in this section. We also provide geometric intuition for how such behavior can be generated using tools like the Marschak-Machina triangles.

\subsection{Worst-Outcome-Shift Problem}
 Recall the preference relation associated with the worst-outcome-shift problem provided in the Introduction where the DM prefers the riskier lottery when both lotteries seem bound for doom/disappointment but prefers the lottery with less risk when a disappointing prize is replaced by a decently valuable prize.

To see why it is not possible to obtain these preferences with the the rank-dependent utility of \textcite{Quiggin1982} or the more general CPT of \textcite{Tversky1992}, note that the worst prize in both lotteries is \$0 at probability 0.9 in both. We merely swapped the common worst component with a prize of \$90 for both lotteries. Therefore, the probability values and the ranks of prizes within a given lottery are unaltered when viewing the \$90 as \$0. For the lottery on the left side, \$200 remains the best prize and \$100 remains the second best in the lottery after the swap, and \$0(\$90) is the worst prize before(after) the swap. The probabilities associated with \$200 and \$100 do not change during this swapping, with \$0/\$90 both at probability 0.9. We have the same type of situation for the lottery on the right side, and so the preferences described in the example cannot occur for the models mentioned.

 Before replacing \$0 with \$90, computing the utility of the lottery on the left side using CPT or rank-dependent utility would give us an expression of the form $w_1 u(\$0)+w_2u(\$100)+w_3u(\$200)$, where $w_1,w_2,w_3$ would denote weights associated with each of the prizes and would be computed in a way specified by the model. Similarly, the utility of the lottery on the right side would be expressed as $w_4u(\$0)+w_5u(\$150)$ for weights $w_4$ and $w_5$. The preference in the example then implies that $$w_1 u(\$0)+w_2u(\$100)+w_3u(\$200)\geq w_4u(\$0)+w_5u(\$150).$$ Given that $\$0$ is the worst prize in both lotteries at probability 0.9 in both, $w_1$ would equal $w_4$. This simplifies the inequality to 
    \begin{equation}
        w_2u(\$100)+w_3u(\$200)\geq w_5u(\$150).
    \end{equation}

Because the probabilities and ranks associated with a prize within each lottery do not change after we swap the \$0 for \$90 in both lotteries, we would use the same weights $w_1,\ldots, w_5$ for the computations of the values of the lotteries. This means that the value of the lottery on the left side becomes $w_1 u(\$90)+w_2u(\$100)+w_3u(\$200)$ while the value of that on the right becomes $w_4u(\$90)+w_5u(\$150).$ The inequality from (1) and the fact that $w_1=w_4$ gives us $$w_1 u(\$90)+w_2u(\$100)+w_3u(\$200)\geq w_4u(\$90)+w_5u(\$150).$$ This implies that it is impossible to have the types of preferences seen in the example where the DM exhibits a change in risk attitude.
    
The idea here is that the weights associated with the prizes when computing the values of lotteries do not change simply because we swap the common worst component with another prize that remains the worst in both lotteries and we do not alter probability values for either lottery. The utility derived from the prizes other than \$0 or \$90 also do not change because there is only one function $u$. As a result, we are forced to prefer the riskier lottery even in a situation where the DM should display risk aversion. 

\begin{figure}[htbp]
    \centering
    \captionsetup[subfigure]{font=footnotesize}

    \begin{subfigure}[t]{0.4\linewidth}
        \centering
        \includegraphics[width=\linewidth]{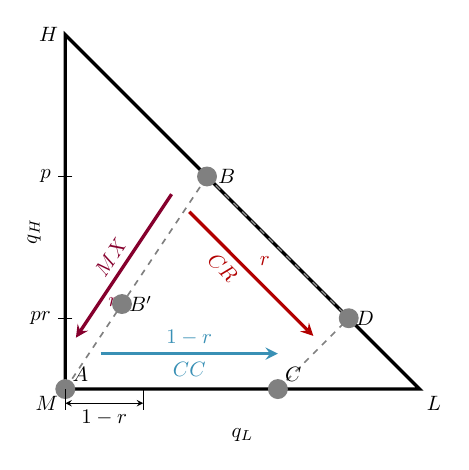}
        \caption{CC/CR/MX geometry}
        \label{fig:crccmx_geometry}
    \end{subfigure}
    \hspace{3em}
    \begin{subfigure}[t]{0.4\linewidth}
        \centering
        \includegraphics[width=\linewidth]{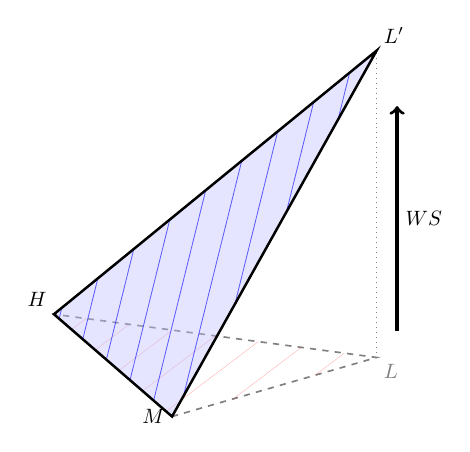}
        \caption{WS}
        \label{fig:ws_illustration}
    \end{subfigure}

    \caption{Visual comparison of CC/CR/MX and the WS.}
    \label{fig:crccmx_vs_ws}
\end{figure}

Figure~\ref{fig:crccmx_vs_ws} provides a visual summary of how the classical
common-consequence (CC), common-ratio (CR), and mixture (MX) comparisons
relate to one another in the Marschak--Machina triangle, and contrasts them with the WS operation studied here. The key distinction is that CC/CR/MX correspond to movements within the probability simplex holding outcome levels fixed, whereas WS shifts a common worst outcome upward while keeping probabilities fixed, leaving rank-dependent decision weights unchanged. Panel \ref{fig:crccmx_geometry} (reproduced from \cite{mcgranaghan2025}) illustrates the standard CC/CR/MX comparisons as movements within the Marschak--Machina triangle with outcome levels held fixed. Panel \ref{fig:ws_illustration} illustrates the WS operation, in which the common worst outcome is increased while all probabilities are held fixed.

To show how the ECU can engender the preference reversal, consider the following example of a binary ECU representation.

    \begin{example}
        Suppose we have a binary ECU where $X=[0,300],d=20,\tau=0.75,$ and we have strictly increasing $u$ such that $u(0)=0,u(90)=15,u(100)=20,u(150)=50,u(200)=60$. Suppose further we have utility function $v$ such that $v(0)=0,v(90)=8,v(100)=10,v(150)=25,v(200)=50$. $\square$ 
    \end{example}

It is easy to verify that the specifications in Example 1 are consistent with strictly increasing $u$ and $v$ and that we have $u(x)>v(x)$ for the values of $x$ that we specified above other than 0. Consequently, we know that a binary ECU with the above specification exists. For the discussion that follows in this section we write an arbitrary lottery $p$ as $p=(x_1,p_1;...;x_n,p_n)$ where, for any $1\leq i\leq n$, $x_i\in X$ and $p_i$ is the probability that $x_i$ is obtained. In some cases, we may have $x_i=x_j$ for $i\neq j$. In this case, denoting this common prize $x$, the probability that $x$ in $p$ is obtained is $\sum_{\lbrace i:x_i=x \rbrace }p_i $. Under this ECU, when the DM evaluates lotteries $p=(0, 0.9; 100, 0.05; 200, 0.05)$ and $q=(0, 0.9; 150, 0.1)$, the utility function $v$ is used because these lotteries place too high a probability on the awarding of disappointing prizes. This gives us 
    \begin{align*}
        V(p)=\int v\; dp=0.9v(0)+0.05v(100)+0.05v(200)=3 \\
        V(q)=\int v\; dq=0.9v(0)+0.1v(150)=2.5.
    \end{align*}
This means that we have $(0, 0.9; 100, 0.05; 200, 0.05) \succ  (0, 0.9; 150, 0.1).$ 
When we take lotteries  $p'=(90, 0.9; 100, 0.05;200, 0.05)$ and $q'=(90,0.9; 150, 0.1)$, however, the DM uses utility function $u$ to evaluate the lotteries because replacing $0$ with $90$ has pushed the probability that disappointments are awarded below threshold $\tau$. Therefore, 
    \begin{align*}
        V(p')=\int u\; dp'=0.9u(90)+0.05u(100)+0.05u(200)=17.5\\
        V(q')=\int u\; dq'=0.9u(90)+0.1u(150)=18.5,
    \end{align*}  
This gives us $(90, 0.9; 100, 0.05; 200, 0.05) \prec  (90, 0.9; 150, 0.1)$, thus generating a change in risk-taking behavior from our initial example.

More generally, to generate preferences of the same type as in the introduction, suppose there is a common worst prize, say $x$, for two lotteries that has been  assigned the same probability weight in both lotteries and is then replaced by a higher prize $y$. We can generate these types of preferences by choosing $d$ to be in $(x,y)$ and choosing $\tau$ to be lower than the probability weight assigned to $x$ in both lotteries.

\subsection{Allais Paradox}
We next turn our attention to preferences associated with the common ratio and common consequence effects of the \textcite{Allais1953} Paradox. We use the specific preferences considered in the experiment conducted by \textcite{Kahneman1979} to consider these two effects. Although the idea seems to carry over from the previous example inasmuch as DMs tend to be risk-loving when disappointment is highly likely but risk-averse when it becomes less likely, the way we modify the lotteries is slightly different from how \textcite{Kahneman1979} did.

For the common consequence effect, we again use a binary ECU to generate the observed preferences.
    \begin{example}
        Suppose we have a binary ECU with $X=[0,3000],d=10,\tau=0.5,$ and a strictly increasing $u$ such that $u(0)=-1000,u(10)=0,u(2400)=2600,u(2500)=2615$. Suppose further that we have a utility function $v$ such that $v(0)=-1000,v(10)=-1,v(2400)=2400,v(2500)=2600$. $\square$ 
    \end{example} 

Take $p=(2500,0.33;0,0.67)$ over $q=(2400,0.34;0,0.66)$. With the specified model, we have $(2500,0.33;0,0.67)\succ (2400,0.34;0,0.66)$, and therefore the DM prefers the riskier lottery of the two when the probability that disappointment occurs exceeds $\tau$. Suppose we replace the $0.66$ probability that $0$ prizes are awarded in both lotteries with $2400$ prizes. The two lotteries, $q'=(2400,1)$ and  $p'=(2500,0.33;2400,0.66;0,0.01)$, are evaluated using $u$ instead of $v$ because the probability that a disappointment occurs vanishes. This gives us preference $q' \succ p'$, and in this case the DM prefers the less risky lottery.

We use a binary ECU also for the common ratio effect.
    \begin{example}
        Suppose we have a binary ECU with $X=[0,7000],d=1000,\tau=0.9,$ and a strictly increasing $u$ such that $u(0)=0,u(1000)=2,u(3000)=25,u(6000)=40$.  Suppose further that we have a strictly increasing utility function $v$ such that $v(0)=0,v(1000)=1,v(3000)=10,v(6000)=30$. $\square$ 
    \end{example}
    \noindent 
With these parameter values, a simple calculation shows that we have $$(6000,0.001;0,0.999)\succ (3000,0.002;0,0.998).$$ Again, the DM prefers the riskier lottery when the probability that  disappointment occurs is high. When we consider lotteries $(3000,0.9;0,0.1)$ and $(6000,0.45;0,0.55)$, where the disappointing prizes are less than or equal to the threshold value, though, we obtain $$(3000,0.9;0,0.1)\succ (6000,0.45;0,0.55),$$ so the DM is risk-averse when considering these lotteries.

\subsection{Violation of the Betweenness Axiom}
\label{sec:betweenness_violation}
    \begin{definition}
        A preference relation $\succsim$ satisfies \textbf{betweenness} if, for any $p,q\in \mathcal{L}$, $p\succ q$ implies $p\succ \alpha p+(1-\alpha)q$ for all $\alpha\in (0,1)$, and $p\sim q$ implies $p\sim \alpha p+(1-\alpha)q$ for all $\alpha\in [0,1]$.  
    \end{definition}
In reference to next set of preferences we discuss, we show how a DM whose risk-taking behavior varies can cause a violation of betweenness. By showing that such preferences can be consistent with the ECU, we demonstrate that our model does not fall into the class of betweenness models. Unlike the examples presented thus far, this example involves a preference for a riskier lottery when the probability that a prize is disappointing is high while a less risky lottery is preferred when the probability is low. 

Consider lottery $p=(\$50,1)$ and lottery $q=(\$100,0.5;\$20,0.5)$. Suppose that the DM considers the 0.5 probability that the lower prize of \$20 is awarded to be disturbingly high and so prefers the less-risky lottery $p$. Now, consider the lottery $$0.6p+(1-0.6)q=(\$100,0.2;\$50,0.6;\$20,0.2).$$ The lower probability that the disappointing \$20 is awarded causes the DM to consider it negligible. This encourages the DM to become risk-loving and take a chance on achieving the highest prize of \$100. This results in $0.6p+(1-0.6)q\succ p$, which is a violation of betweenness.

    \begin{example}
        Suppose we have a binary ECU. Let $X=[0,200],d=30,\tau=0.3,$ with a strictly increasing $u$ such that $u(20)=8, u(50)=10, u(100)=20$. Suppose further that we have a utility function $v$ such that $v(20)=4, v(100)=12$. $\square$ 
    \end{example} 

Lottery $p$ places probability 0 on disappointing prizes so its value based on the ECU representation is $V(p)=1u(50)=10$. Lottery $q$, on the other hand, assigns a probability higher than $\tau=0.3$ to prizes that are less than or equal to $d=30$, so its value is given by $V(q)=0.5v(100)+0.5v(20)=8$. This gives us $p\succ q$. In contrast, lottery $0.6p+(1-0.6)q$ assigns only a 0.2 probability to the \$20 prize, and this means that the lottery will be evaluated according to $u$. We thus have $V(0.6p+(1-0.6)q)=0.2u(100)+0.6u(50)+0.2u(20)=11.6$, which implies that $0.6p+0.4q\succ p$. The intuition here is that the DM uses function $v$ when evaluating lottery $q$ because the high probability of disappointment makes the DM pessimistic. The lower probability that  disappointment occurs when we take the mixture $0.6p+0.4q$, however, makes the DM feel optimistic and leads the DM to use function $u$ when evaluating the lottery.

\subsection{Violation of Negative Certainty Independence}
The cautious expected utility model of \textcite{Cerreia2015} is closely related to our ECU since both models allow violations of independence and involve multiple utility functions. The key axiom  which serves as one of the necessary and sufficient conditions for the cautious expected utility model is negative certainty independence. 

\begin{definition}
    A preference relation $\succsim$ satisfies \textbf{negative certainty independence (NCI)} if, for any $p,q\in\mathcal{L}$, any $x\in [w,b]$, and any $\alpha\in[0,1]$,
    \[
    p \succsim \delta_x \ \implies\ \alpha p + (1-\alpha)q \succsim \alpha \delta_x + (1-\alpha)q .
    \]
\end{definition}

The following example shows that there are specifications of the ECU that are not consistent with the cautious expected utility. We do so by showing that the ECU can exhibit violation of negative certainty independence. As in all of the other examples, the idea is that the DM exhibits different risk attitudes depending on the context of the lottery.

Consider the lottery $p=(\$35,0.8;\$100,0.2)$. The DM considers the 0.2 probability of getting the high prize of $\$100$ to be significant enough and so prefers this more risky lottery over the certainty equivalent $\delta_{48}=(\$48,1)$. However, when considering the lottery $(\$0,0.9;\$35,0.08;\$100,0.02)$, the probability of getting $\$100$ is so low that instead of taking a chance at getting  $\$100$, it prefers the lottery $(\$0,0.9;\$48,0.1)$. Any standard EU maximizer who is risk-averse would prefer the latter over the former. Therefore, the DM seems to be risk-loving when comparing the first pair of lotteries but exhibits risk-aversion for the second pair.

The described preferences is a violation of the Negative Certainty Independence. To see this, if we let $q=(\$0,1)$ and $\alpha=0.1$, the second pair of lotteries is $\alpha p+(1-\alpha)q$ and $\alpha \delta_{48}+(1-\alpha)q$. In this example, we have $p\succ \delta_{48}$ but $\alpha p+(1-\alpha)q\prec \alpha \delta_{48}+(1-\alpha)q$.

\begin{example}
	Let $X=[0,100]$. Suppose we have a Binary ECU with $\tau=0.5$ and $d=20$. Define functions $u$ and $v$ such that $v(0)=u(0)=0,v(100)=u(100)=10, u(35)=4,u(48)=5,v(35)=1,v(48)=3$. It is easy to see that there exist functions $v$ and $u$ that are consistent with these values and satisfy the conditions for being a Binary ECU. The preferences represented by such a Binary ECU would have $p\succ \delta_{48}$ and $\alpha p+(1-\alpha)q\prec \alpha \delta_{48}+(1-\alpha)q$.
\end{example}

\subsection{Illustrating Changes in Risk Attitudes}
To provide a clearer intuition for how the model gives rise to all the mentioned behaviors, we illustrate the familiar Marschak-Machina triangles for the binary ECU. As usual, we let the top left corner represent the largest prize (H) among the three vertices, while the bottom left corner represents the medium value (M) and the bottom right corner represents the lowest value (L). Whenever all three prizes are greater than d ($H>M>L>d$), the MM-Triangle is identical to that of the standard EU. This is also the case if $d>H>M>L$. Thus, the only interesting cases are $H>M>d>L$ and $H>d>M>L$. We present these two cases in that order. 

\begin{figure}[ht]
    \centering
    
    \begin{subfigure}[b]{0.4\textwidth}
        \centering
        \includegraphics[width=\textwidth]{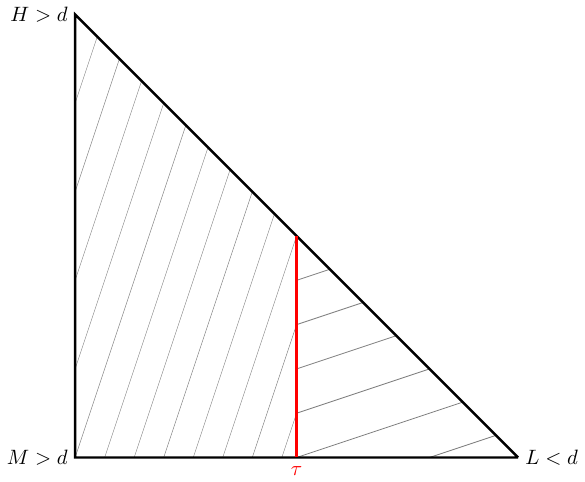}
        \caption{$H>M>d>L$}
        \label{fig:mmt_1}
    \end{subfigure}
    \hspace{2em}
    \begin{subfigure}[b]{0.4\textwidth}
        \centering
        \includegraphics[width=\textwidth]{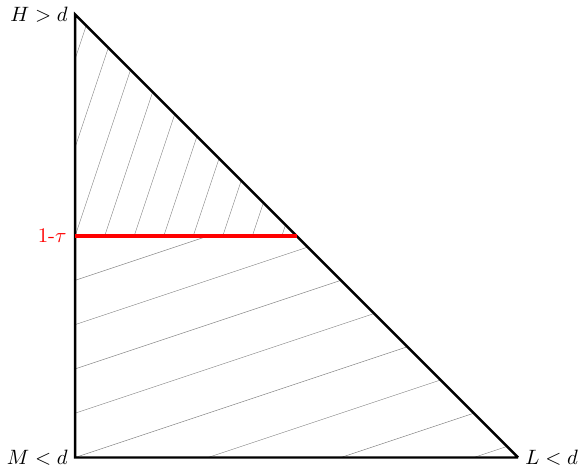}
        \caption{$H>d>M>L$}
        \label{fig:mmt_2}
    \end{subfigure}
    
    \caption{Marschak-Machina triangles under different threshold orderings.}
    \label{fig:mmt_combined}
    
\end{figure}

For this first case, because L is the only prize below $d$, we have the usual EU maximizer using function $u$ as long as the probability placed on L is less than or equal to $\tau$. Once the probability rises above $\tau$ at the red line, the DM switches to function $v$ and this can generate the change in slopes of the indifference  curves. So, instead of a gradual \enquote{fanning out}, we see a sudden change in slopes when we leave the bottom right region of the triangle.

The idea informing the second case is similar. When H is the only prize above the threshold, we have a standard EU maximizer using function $v$ whenever the probabilities placed on M and L exceed $\tau$. This means the probability placed on H comes in below $1-\tau$. When the probability of H exceeds $1-\tau$ at the red line, we obtain the change in functions from $v$ to $u$.

    \begin{figure}[ht]
        \centering
        \includegraphics[width=0.5\linewidth]{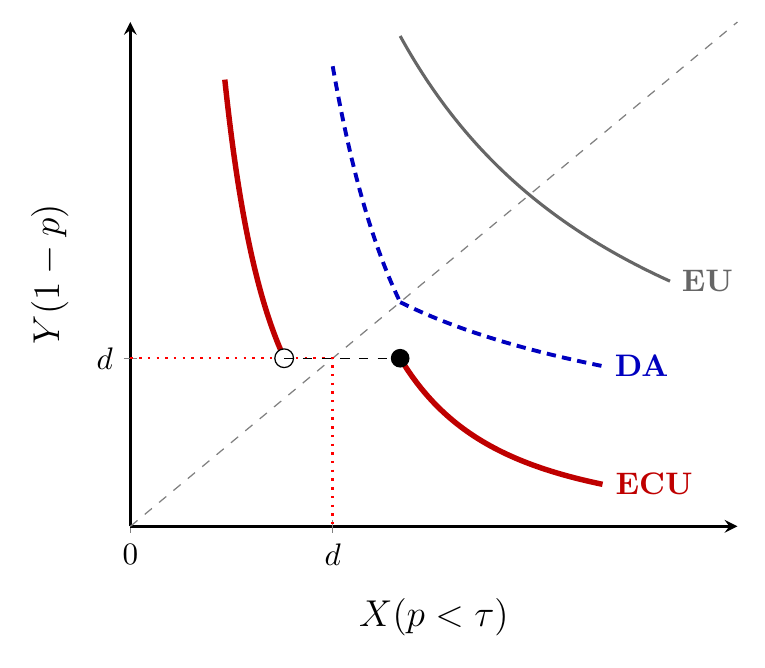}
        \caption{Visual Comparison: EU, DA, and ECU}
        \label{fig:ecu-da-eu-comparison}
    \end{figure}

Additionally, we provide the types of indifference curves given in Figure \ref{fig:ecu-da-eu-comparison} of \textcite{Gul1991} using another binary ECU example. For this graph, we let the horizontal axis represent the value of a prize $X$ that occurs in a lottery at probability $p$ and the vertical axis represent the value of a prize $Y$ that occurs in the same lottery at probability $1-p$. So point $(x,y)$ in our diagram corresponds to a lottery that awards prize $x$ at probability $p$ and prize $y$ at probability $1-p$. We focus on the case in which $p<\tau <0.5$. Each indifference curve corresponds to lotteries that are indifferent to each other. For ease of comparison, we draw these curves for the case in which we obtain smooth indifference curves when the DM is a standard EU maximizer.

For Figure \ref{fig:ecu-da-eu-comparison}, in the bottom indifference curve we initially see the usual smooth curve until Y drops to value $d$ because we have a standard EU maximizer using function $u$ as a result of $p<\tau$. Once the value of Y reaches $d$, prize Y is disappointing and carries probability $1-p > \tau$, so the DM switches to function $v$. When this happens, the utility that the DM derives from prize Y suddenly drops from $u(Y)$ to $v(Y)$. To keep the DM indifferent, the value of X needs to rise much higher to compensate for this change, and hence there is a discontinuity at that point. As the higher indifference curve shows, we have a smooth indifference curve without any discontinuity whenever the probability placed on prizes that are less than or equal to $d$ is below $\tau$. We obtain this result because, of course, the DM will always use $u$ to evaluate such lotteries.

\section{Representation Theorem and Uniqueness}
This section presents some of the main theoretical results. These include the axiomatic characterization of the representation along with the uniqueness of the model ingredients.
\subsection{Axioms}
We now present the axioms that are necessary and sufficient for an ECU representation. In what follows, we may describe a lottery $p$ as $p=(x_1,p_1;...;x_n,p_n)$ where, for any $1\leq i\leq n$, $x_i\in X$ and $p_i$ is the probability that $x_i$ is obtained. In some cases, we may have $x_i=x_j$ for $i\neq j$. In this case, denoting this common prize as $x$, the probability that $x$ is obtained in $p$ is $\sum_{\lbrace i:x_i=x \rbrace }p_i $. The first several axioms are straightforward.
 
    \begin{axiom}
        $\succsim$ is complete and transitive.
    \end{axiom}
    \noindent 
We let $\succ$ and $\sim$ denote the asymmetric and symmetric components of $\succsim$, respectively. 

    \begin{axiom}[\textbf{Monotonicity}]
        $\forall t_1,t_2\in [0,1]$, $t_1\delta_b +(1-t_1)\delta_w\succsim t_2\delta_b+(1-t_2)\delta_w$  iff $t_1\geq t_2$. 
    \end{axiom} 

Axiom 2 is a natural restriction which says that, when we have two lotteries where both are combinations of the worst and best prizes, the one that places greater weight on the best prize is preferred.

    \begin{axiom}[\textbf{Replacement Monotonicity}]
        For any $x\in X$ and $\alpha\in [0,1]$, we have $$\alpha\delta_b+(1-\alpha)\delta_x\succsim \alpha\delta_b+(1-\alpha)\delta_w\succsim \alpha\delta_x+(1-\alpha)\delta_w.$$
    \end{axiom} 

To understand this axiom, consider first the middle lottery in the expression of the axiom, which is a combination of the best and worst prizes. The first preference relation says that, if we replace the worst prize with any other $x$, then the resulting lottery is preferred over the middle lottery. Similarly, the second preference relation says that, if we replace the best prize $b$ with any other $x$, then this makes the new lottery less preferable than the middle lottery. We call this the replacement monotonicity axiom for this reason. 
 
    \begin{axiom}[\textbf{Weak Solvability}]
        For any $p\in \mathcal{L}$, there exists $\gamma\in [0,1]$ such that $p\sim \gamma \delta_b+(1-\gamma )\delta_w$.
    \end{axiom}

Axiom 4 states that any lottery can be made indifferent to a combination of the worst and best prizes. It resembles the solvability of \textcite{Dekel1986}, but we restrict attention to combinations of the best and worst prizes only. As such, we call this the weak solvability axiom. 

Given weak solvability, we can define a value $\phi_x$ for each $x\in X$ where $\delta_x\sim \phi_x \delta_b+(1-\phi_x)\delta_w$. Note that this value is unique because of Axiom 2. We use the values $\phi_x$ to define the set $${\mathcal{D}}=\left \lbrace x\in [w,b):\alpha\delta_x+(1-\alpha)\delta_w \sim \alpha \phi_x \delta_b+\alpha(1-\phi_x)\delta_w+(1-\alpha)\delta_w\text{ }\forall \alpha\in [0,1] \right \rbrace.$$ $w$ is included in the set, so $\mathcal{D}\neq \emptyset$. Therefore, we let $\tilde{d}=\sup \mathcal{D}$.  It turns out that $\tilde{d}$ equals the threshold value $d$ if $\succsim$ has an ECU representation and hence can be used as the threshold value for disappointing prizes when constructing an ECU representation from preferences. This ends up being true because ${\mathcal{D}}=[w,d]$ when $d\neq b$ and ${\mathcal{D}}=[w,b)$ when $d=b$. 

The intuition that explains why this is true is quite simple. First, consider the case in which $d\neq b$. If $x\in [w,d]$, then lottery $\alpha\delta_x+(1-\alpha)\delta_w $ on the left side of $\sim$ in the definition of $\mathcal{D}$ would place probability 1 on disappointing prizes. As such, the lottery would be evaluated as $\alpha u_1(x)+(1-\alpha)u_1(w)$. We can rewrite the first part as a convex combination of the utilities of the best and worst prizes to obtain $$\alpha \phi_x u(b)+\alpha(1-\phi_x)u(w)+(1-\alpha)u(w).$$ As this would be the EU of lottery $\alpha \phi_x \delta_b+\alpha(1-\phi_x)\delta_w+(1-\alpha)\delta_w$, we obtain the indifference relation as in the definition of $\mathcal{D}$. Note that this argument is independent of the value of $\alpha$ and allows us to conclude $x\in \mathcal{D}$.

For any $x\in (d,b)$, it will be the case that $x\notin \mathcal{D}$. This tells us that ${\mathcal{D}}=[w,d]$. As a result, we can take the supremum of the set to pin down the threshold value. To see why this is the case, suppose that $x\in (d,b)$. We then have $V(\delta_x)=u_0(x)$, so the value $\phi_x$ such that $\delta_x\sim \phi_x\delta_b+(1-\phi_x)\delta_w$ can be found by solving equation $u_0(x)=\phi_x u(b)+(1-\phi_x) u(w)$. Because $d\neq b$, by assumption of an ECU representation there exists an $\alpha\in (0,1)$ such that $u_\alpha(x)\neq u_0(x)$. Thus, if we take  $p=(1-\alpha)\delta_x+\alpha\delta_w$, we have 
    \begin{align*} 
        V(p)&=(1-\alpha)u_\alpha(x)+\alpha u_\alpha(w)\\
        &\neq (1-\alpha)u_0(x)+\alpha u_\alpha(w)\\
        &=(1-\alpha) \phi_x u(b)+(1-\alpha)(1-\phi_x)u(w)+\alpha u(w)\\
        &=V((1-\alpha)\phi_x\delta_b+(1-\alpha)(1-\phi_x)\delta_w+\alpha  \delta_w).
    \end{align*}
This implies that we do not have $(1-\alpha)\delta_x+\alpha\delta_w\sim (1-\alpha)\phi_x\delta_b+(1-\alpha)(1-\phi_x)\delta_w+\alpha  \delta_w$, which signifies $x\notin \mathcal{D}$.

For the case where $d=b$, if we take any $x\in [w,b)$, we can use the same argument for the case of $d\neq b$ and $x\in [w,d]$ to show that $x\in \mathcal{D}$. This gives us ${\mathcal{D}}=[w,b)$. We use the value $\tilde{d}$ to define our final axiom, which we introduce after providing a lemma.

Axioms 1-4 allow us to prove the following lemma, which states that, if we have a lottery consisting of a convex combination of some prize $x$ with either $w$ or $b$, so that we have $\alpha\delta_x+(1-\alpha)\delta_y$ where $y$ is $b$ or $w$, we can replace the component $\alpha\delta_x$ with some combination of the best and worst prize and maintain indifference. Thus, we can think of this lemma as the statement that we have solvability of the component $\alpha\delta_x$ with respect to $\delta_w$ and $\delta_b$.

    \begin{lemma}
        \label{lemma:component-solvability}
        For any $x\in X, y\in\{w,b\},\alpha\in[0,1]$, there exists $\gamma\in [0,1]$ such that $\alpha\delta_x+(1-\alpha)\delta_y\sim \alpha\gamma\delta_b+\alpha(1-\gamma) \delta_w+(1-\alpha)\delta_y$
    \end{lemma} 
    \begin{proof}
        See Appendix \ref{appendix:proofs}.
    \end{proof}

With this lemma in hand, we move on to define several values we need to formulate the final axiom. For any disappointing prize $x\in[w,\tilde{d}]$ and $\alpha\in[0,1]$, we define the value $\phi^\alpha_x$ by taking lottery $\alpha\delta_x+(1-\alpha)\delta_b$ and letting 
$$\alpha\delta_x+(1-\alpha)\delta_b\sim \alpha\phi_x^\alpha\delta_b+\alpha(1-\phi^\alpha_x)\delta_w+(1-\alpha)\delta_b .$$ Notice that there is a probability $\alpha$ that prize $x$ is won in the lottery on the left side. We are simply replacing that prize $x$ with a combination of $b$ and $w$, with the weight on $b$ being $\phi_x^\alpha$ to obtain the lottery on the right. Thus, the value $\phi_x^\alpha$ tells us the weight we need on $b$ to obtain indifference. These values are well-defined by Lemma \ref{lemma:component-solvability}.

To understand why we define these values as we do, it is useful to think about how the standard EU model works. In the standard model, if we were to take any prize $x\in [w,b]$, we could make it indifferent to the convex combination of the best and worst prizes by finding a value of $\phi_x$ such that $u(x)=\phi_xu(b)+(1-\phi_x)u(w)$. The value $\phi_x$ is the weight that needs to be placed on the best prize. Now, given any lottery $p$ with $x\in supp(p)$, we can generate a lottery that is indifferent to $p$ by replacing $x$ with a convex combination of best and worst prizes with weight $\phi(x)$ on $b$. For example, if $p=(w,0.2;x;0.8)$ then
    \begin{align*}
        V(p)&=0.2u(w)+0.8u(x)\\
        &=0.2u(w)+0.8\phi_xu(b)+0.8(1-\phi_x)u(w),\\
    \end{align*}
so that $p\sim (w,0.2+0.8(1-\phi_x);b,0.8\phi_x)$. This helps us pin down the utility values when constructing an EU representation, usually by setting $u(x)=\phi_x$.

We need a similar condition for our model. That is, we want to be able to replace a prize $x$ with a combination of $b$ and $w$. There is however a slight complication. Note that the value of $\phi_x$ in the argument for the standard EU model is independent of lottery $p$. More precisely, it is independent of the probability value that a lottery assigns to disappointing prizes. This is not the case for our model because the utility function used to evaluate a lottery depends on the probability that prizes have disappointing values. When $p([w,d])=\alpha$, the utility derived in this lottery by prize $x$ may be $u_\alpha(x)$, such that $u_\alpha(x)=\phi u(b)+(1-\phi)u(w)$ for some $\phi$ in the ECU. 

For $q\in \mathcal{L}$ such that $q([w,d])=\beta\neq \alpha$, however, the utility from prize $x$ may be $u_\beta(x)\neq u_\alpha(x)$. In this case, we will not have $u_\beta(x)=\phi u(b)+(1-\phi)u(w)$ if we use the same $\phi$ as above. Therefore, we need dependence on $\alpha$ and require the use of $\phi_x^\alpha$. The method for finding this value depends on whether $x\in [w,\tilde{d}]$ or not. 

First, for $x\in[w,\tilde{d}]$ and $\alpha\in (0,1]$, we need lottery $\alpha\delta_x+(1-\alpha)\delta_b$ because it places probability $\alpha$ on disappointing prizes (unless $b$ is also considered disappointing, but we do not need to vary $\alpha$ in this case as this reduces ours to a standard EU model) and the value in the ECU would be computed according to 
$$\alpha u_\alpha(x)+(1-\alpha)u(b).$$ 

We let $\phi_x^\alpha$ be the value such that $$\alpha\delta_x+(1-\alpha)\delta_b\sim \alpha\phi_x^\alpha\delta_b+\alpha(1-\phi^\alpha_x)\delta_w+(1-\alpha)\delta_b .$$ We would then be able to use $\phi_x^\alpha$ as $u_\alpha(x)$. 

We need similar values for $x\notin [w,\tilde{d}]$ and $\alpha\in[0,1)$. In these cases, we take lottery $\alpha\delta_w+(1-\alpha)\delta_x$ and let $\phi_x^\alpha$ be the value such that $$\alpha\delta_w+(1-\alpha)\delta_x\sim\alpha\delta_w+(1-\alpha)\phi_x^\alpha\delta_b+(1-\alpha)(1-\phi_x^\alpha)\delta_w.$$ The idea again is that we are replacing $x$ with a combination of $b$ and $w$ with weight $\phi_x^\alpha$ on $b$. Because $x$ is not a disappointing prize, lottery $\alpha\delta_w+(1-\alpha)\delta_x $ places probability $\alpha$ on disappointing prizes. 

We can use these values to state the final axiom.This axiom says simply that, if we have a lottery that places probability $\alpha$ on disappointing prizes, we can substitute each prize in the support with the \enquote{right combinations} of $b$ and $w$ to maintain indifference. We say \enquote{right combinations} because the weights depend on the lottery's context, which is the probability placed on what seem to be disappointing prizes.

    \begin{axiom}[\textbf{Contextual Substitutability}]
        \label{axiom:contextual_substitutability}
        For any $\alpha \in [0,1]$ and $p\in \mathcal{L}$ such that $p([w,\tilde{d}])=\alpha$, $$p\sim \sum_{x\in supp(p)}p(x)[\phi_x^\alpha\delta_b+(1-\phi_x^\alpha)\delta_w].$$
    \end{axiom} 

Finally, although we do not state it as a formal axiom since it is just a mild regularity condition that ensures we have a set of contextual utility functions, we need the following condition. The condition along with the other axioms are necessary and sufficeint for a ECU representation. 
\begin{RC}
If $\tilde{d}\neq b$, then for any $x\in (w,b)$ there exists $\mu,\pi\in(0,1)$ such that $\phi_x^\pi\neq \phi_x^\mu$.
\end{RC}

    \begin{theorem}
        Preference relation $\succsim$ has an ECU representation iff it satisfies completeness, transitivity, monotonicity, replacement monotonicity, weak solvability, contextual substitutability, and the regularity condition.
    \end{theorem}

    \begin{proof}
        See Appendix \ref{appendix:proofs}.
    \end{proof}

\subsection{Interpretation of Contextual Substitutability}
\label{sec:axiom_interpretation}
The contextual substitutability axiom appears rather technical, but the underlying idea is simple and admits a natural behavioral interpretation. It says that the way the DM values a prize is consistent across lotteries as long as there is no change in the context. So, for any lottery $p$ and $q$ that have the same probability $\alpha$ of disappointment, we can measure the value of any prize $x$ to a DM in this context with $\phi_x^\alpha \delta_b+(1-\phi_x^\alpha)\delta_w$. To see this idea clearer, we think about the following axiom which is a straightforward weakening of the usual independence axiom. It states that independence is satisfied for combinations of lotteries that have the same context.
 
 \begin{axiom}[Contextual Independence]
    \label{axiom:contextual_independence}
 	For any $\gamma\in [0,1]$ and $p,q,r\in \cal{L}$ such that $p([w,d])=q([w,d])=r([w,d])$, $p\succsim q\implies \gamma p+(1-\gamma)r\succsim \gamma q+(1-\gamma)r$. 
 \end{axiom}

The condition $p([w,d])=q([w,d])=r([w,d])$ tells us that all three lotteries have the same context, meaning the probability of disappointment. Needless to say, the usual Independence axiom places the restriction on all combinations of lotteries $p,q,r$ and not just for ones with the same context. Since the idea of Contextual Substitutability is that the way the DM values prizes are consistent when the context is unaltered, intuition tells us that this version of Independence should be implied by the ECU even if the standard independence may be violated.

We note a deeper structural point here. Axiom \ref{axiom:contextual_substitutability} plays two roles simultaneously: it requires that prizes combine linearly with probabilities within any fixed context, and it requires that the substitution values $\phi_x^\alpha$ across different contexts are calibrated accordingly so that prize-for-mixture replacements preserve indifference regardless of which context the lottery belongs to. Contextual Independence captures only the first requirement. It is silent on how evaluations should relate across contexts with different disappointment probabilities.

When the context dimension is trivial, i.e., when $\tilde{d}=b$ so that every lottery has the same disappointment probability, this distinction vanishes. There is only one context, so cross-context calibration is vacuous. The two requirements that Independence enforces are logically fused precisely because there is no variation along the context dimension to separate them. Once $\tilde{d}\neq b$ introduces a nontrivial context dimension, the fusion breaks. The next proposition shows Contextual Independence remains necessary but is now strictly weaker than what Axiom~5 demands because of this reason. In this sense, the standard Independence axiom was not a sufficiently foundational notion to render behavioral consistency; it only appeared to be one because the context dimension was degenerate.

\begin{proposition}
    \label{prop:contextual_independence}
	If $\succsim$ is an ECU, then it satisfies contextual independence.
\end{proposition}

While less primitive than Contextual Subsitutability in a mechanical sense, Contextual Independence bears significance for the following reason. The ECU may seem as being too flexible and allowing any type of behavior so that it lacks a meaningful testable implication. This proposition says otherwise and provides a restrictive testable implication that is easy to verify. For example, in an experimental setting inside a lab, once threshold value $d$ is recovered from the preferences of a DM, the experimenter can verify whether the axiom is violated or not by mixing lotteries with the same contexts. Equivalently, any observed violation of Independence must involve
a change in the common context: if mixing preserves the disappointment
probability, the ranking is preserved.

To elaborate, suppose we take two lotteries $p$ and $q$ that have small probabilities of disappointment and have the same context, meaning that $p([w,d])=q([w,d])$. Since the lotteries have small disappointment probabilities the utility function being used to evaluate them could be a concave one representing a risk-averse attitude. If we have $p\succsim q$, then the proposition tells us that we cannot have $\gamma q+(1-\gamma)r \succ \gamma p+(1-\gamma)r$ if the lottery $r$ also has the same context with $r([w,d])=p([w,d])$. This is because mixing in the lottery $r$ to $p$ and $q$ does not change the context of the decision problem since $\gamma p+(1-\gamma)r$ and $\gamma q+(1-\gamma)r$ also have the same disappointment probabilities  as $p,q,r$. However, if $r$ is a lottery with an extremely high level of disappointment, the mixtures may end up having a high probability of disappointment. This change in the context of the lotteries can thus cause the DM to switch to a risk-loving utility function and generate a violation of independence. Proposition~\ref{prop:contextual_independence} therefore yields a simple falsification strategy that does not require structural estimation: violations can only arise when mixing changes the common context.

\subsection{Uniqueness}
We now turn our attention to whether the model ingredients can be identified to some extent. This means that we want to know whether a given preference relation consistent with the ECU must have a unique set of model ingredients that generate the preference relation. Recall that for the standard EU, the utility function over prizes can be identified to the extent that all utility functions that generate a given preference relation must be positive linear transformations of one another. The hope is that a similar result holds for the ECU. It turns out that such a sharp uniqueness result also applies for the ECU. Namely, $d$ is unique and all the functions are unique up to positive linear transformations as long as we do not have a preference relation consistent with $d=b$. We can even assert that the constants for the linear transformations must be the same across all utility functions.

\begin{theorem}
\label{thm:uniqueness}
	Suppose preference relation $\succsim$ has ECU representations with $\left \lbrace d_1,(u_\alpha)_{\alpha\in[0,1]}\right \rbrace $ where $d_1<b$. Then $\left \lbrace d_2,(v_\alpha)_{\alpha\in[0,1]}\right \rbrace $ is another ECU representation of $\succsim$ iff $d_1=d_2$ and $v_\alpha=ku_\alpha+c$ for $k\in \mathbb{R}_{++}$ and $c\in \mathbb{R}$ for all $\alpha\in[0,1]$.
\end{theorem}
\begin{proof}
	See Appendix \ref{appendix:proofs}.
\end{proof}

We need preferences where we do not have $d=b$ for this theorem since those would be the preferences for a standard EU maximizer. In such a case, the DM would evaluate all lotteries using the function $u_1$. Therefore, all other utility functions become irrelevant and we cannot guarantee any sort of uniqueness for the other utility functions $u_\alpha$ for $\alpha\neq 1$. Even for such a preference relation of a standard EU maximizer, we point out that $d$ is unique and the relevant function $u_1$ is unique up to a positive linear transformation. The meaningful ingredients are thus unique for this case as well.

We want to emphasize that a DM may utilize only a small number of distinct utility functions. This also means that, from a practical standpoint, observed choice data can be rationalized with a small number of distinct functions. For example, in the case with the binary ECU, all $u_\pi$ for $\pi < \tau$ are identical, and vice versa. To get a sense of the size of the set of different utility functions being used, we can partition the set $[0,1]$ by letting values in the same equivalence class be the disappointment probability values that use the same utility function. That means $\pi$ and $\mu$ are in the same equivalence class if $u_\pi=u_\mu$. Each equivalence class would then correspond to a distinct utility function. 

One may be concerned that the equivalence classes generated in this way may not be the same for two ECUs that represent the same preference relation so that the the number of distinct utility functions being used by the DM may not be identifiable. An important implication of Theorem \ref{thm:uniqueness} is that all ECUs that represent the same preference relation must have the same partition except for the trivial case of $d=b$. This result assures we do not face a situation where the same preference relation could be represented by different numbers of distinct utility functions. 

To formalize the notion that the partition is unique, consider an ECU with utility functions $\{u_\alpha\}_{\alpha\in [0,1]}$. We define a binary relation $\overset{u}{\sim}$ over $[0,1]$ as follows. For any $\pi,\mu\in [0,1]$, let $\pi\overset{u}{\sim} \mu\iff u_\pi=u_\mu$. The relation $\overset{u}{\sim}$ is clearly reflexive, symmetric, and transitive. Therefore, $\overset{u}{\sim}$ is an equivalence relation and the equivalence classes form a partition. We state the result as a corollary since the proof is trivial after Theorem \ref{thm:uniqueness} is established.

\begin{corollary}
Suppose a preference relation $\succsim$ has an ECU representation with utility functions $\{u_\alpha\}_{\alpha\in[0,1]}$ and $d\neq b$. If $\succsim $ has another ECU representation with utility functions $\{v_\alpha\}_{\alpha\in[0,1]}$, then $\overset{u}{\sim}=\overset{v}{\sim}$.
\end{corollary}

We need the threshold $d$ to be strictly less than $b$ in the corollary for the same reason we needed the condition in Theorem \ref{thm:uniqueness}. For the case with $d=b$, we have a preference relation of a standard EU maximizer and we cannot get a uniqueness result because all functions other than $u_1$ are irrelevant.

\section{Empirical Evidence: Worst-Outcome-Shift}
\subsection{Design and procedures}
We conducted a simple experiment to document WS patterns using lotteries displayed in a multiple price list (MPL) interface following the tradition of \textcite{Holt2002}, \textcite{Chapman2017}, and, more recently, \textcite{Bernheim2020}. Each row of the table represent a task, which is our basic unit of observation. Each task asks participants to choose between “Option A” and “Option B.” Option A is a two-outcome lottery and Option B is a three-outcome lottery, much as in the original Allais example. Prizes are denominated in points with each point later converted to \$0.01 USD. Tasks are grouped into three stages. The first two stages are most central to our study of the WS problem. The third stage adds additional tasks and serves as a robustness check. We summarize its role briefly below. We include screenshots of the experimental software interface in Appendix \ref{app:screenshots} and the full experimental instructions in Appendix \ref{app:instructions}.

In Stage 1, we present a table with ten rows, where each row displays a task with Option A on the left column, Option B on the right, and arrows in the middle to indicate the participant's choice between the options. We constructed the prizes using the same ratio as in the motivating example in the introduction. Across all rows, Option A offers a two-outcome lottery and Option B offers a three-outcome lottery. Both lotteries include the same worst outcome with the same probability, as in a canonical common-consequence problem. Within each option, probabilities are constant across rows. All prizes are also fixed across all rows, except the worst outcome common in both lotteries. We increase the worst outcomes in increments of 44 points over the rows, from 0 to 400 points. We additionally set expected values of Option A equal to that of Option B in all ten tasks. 

As the worst outcome enters both options with the same probability and the remaining outcomes and probabilities are held fixed across rows, EU and rank-dependent models predict invariance in this common-consequence design; thus any switching across rows violates that invariance. This design feature parallels the rank-independence evidence emphasized by  
\textcite{Bernheim2020}. Note that the first and the last rows already constitute a pair that is analogous to the original Allais experiment. Thus, a choice reversal between the two tasks is sufficient to violate EU and rank-dependent models, as argued previously. The intermediate rows provide a finer grid to understand the choices with respect to the context of lotteries. Also, under the specific choices of parameters in the setup, the disappointment aversion model predicts no switching, as the worst outcome remains below the lotteries' respective certainty equivalence across all rows; thus, any observed switching is not consistent with DA. Appendix \ref{app:ws-non-eu} shows the computation using the parameters of our experiment setup.

The interpretation of choice reversals across ten rows may, however, be more complicated than the Allais reversal with two tasks. Potential problems that may arise when participants switch between columns more than once include irregular switching (\textcite{Chew2022}) and a preference for randomization (\textcite{Agranov2017, Feldman2022}). While multiple switching may genuinely reflect richer structure in preferences that is not captured by standard models, our main objective is to document the WS regularity, not to identify the source of such behavior. Restricting attention to single-switch behavior in Stage 1 where we only vary the size of the worst outcome lets us use the switching location as a coarse interval estimate of \( d \) under mild monotonicity assumptions.

For this reason, we restricted participants to making only one switch in the main sessions for more straightforward interpretability of data\footnote{We allowed participants to switch freely across rows in the first two pilot sessions. We report them in Appendix~\ref{app:unrestricted-switching}.
 We return to the multiple switching behavior as a future direction in Section~\ref{sec:remark}.}. Participants were instructed to draw a line on the table, indicating the row at which they would switch from preferring Option A to preferring Option B (or vice versa), rather than making independent decisions in each row. Participants can move a single cutoff up or down by clicking one of the arrows displayed in the center column of the table. All rows above the cutoff were assigned to one option and all rows below to the other. This restriction allows us to interpret the switching point with respect to the payoff threshold more easily. Assuming a binary ECU, a switch between Task \(k\) and Task \(k+1\) implies that the threshold \(d\) falls between the worst outcome in  Task \(k\) and the worst outcome in Task \(k+1\): \( d \in (44 \times k, 44 \times (k+1))\).

We present another table of ten rows in Stage 2. Unlike Stage 1, here we fix the prizes but vary the probabilities placed on the disappointing outcome. More specifically, we construct lotteries with hypothetically not-disappointing prizes above individual specific threshold \(d\), using the Stage 1 switching location as its coarse estimate\footnote{For participants who do not switch within the 10-row range, we code the implied threshold as lying above the design grid by assigning it to the upper endpoint plus one increment; we apply the same convention when constructing \( \tau \) after Stage 2.}. The lotteries we used in the experiment has the following format:
	\begin{equation*}
	 	\left( d+ \frac{b-d}{2}, 1-y; 0, y \right) \text{ vs. } \left( d + \frac{3b-3d}{4}, \frac{1-y}{2}; \quad d + \frac{b-d}{4}, \frac{1-y}{2}; 0, y \right),
	 \end{equation*}
 where \(d\) is recorded as \(44 \times (k+1)\) when a participant switches between Task \(k\) and Task \(k+1\)\footnote{We code \(d \) using the upper grid point of the interval to be conservative when constructing Stage-2 lotteries, ensuring that the non-disappointing prizes are certainly above the entire inferred interval.}, and \( b \) denotes the upper prize level used in the experiment. The left-side lottery places probability \( y \) on the supposedly disappointing outcome 0 and \( 1-y \) to the non-disappointing outcome \( d + (b-d)/2\). The right-side lottery splits the probability \( 1-y \) assigned to two non-disappointing outcomes that are chosen symmetrically around the left-side non-disappointing outcome. This mean-preserving split makes the right-side lottery riskier.

With this construction, we fix the worst outcome at 0 and change the respective probability \(y\) in fixed increments from 0 to 90\% across the ten rows. As in Stage 1, here we can interpret such a switch as a demonstration of a scenario in which a participant uses one utility function for disappointing outcome 0 as too likely and another utility function for the same disappointing outcome 0 as less likely. Note that a switch between the two columns in Stage 2 violates EU but not necessarily rank-dependent utility/CPT, as we vary probabilities across tasks. Again, restricting participants to making only one switch and assuming binary ECU, we interpret the switching point from one option to another as the tolerance threshold \( \tau \).

The main idea behind Stage 3 is a robustness check of replicating standard Allais-like behavior using ECU framework. As shown in Figures \ref{fig:mmt_1} and \ref{fig:mmt_2}, the DM in our ECU model will change their risk attitudes when crossing threshold lines on their Marschak-Machina triangles. That is, we should expect to see an Allais-like reversal if we give them choice tasks for lotteries on opposite sides of a threshold line. On the other hand, we should not expect to observe such a reversal if we provide lotteries only on one partition of the triangle. Using the Stage 1 and Stage 2 switching locations in place of \(d\) and \( \tau \) in the binary ECU model, we construct common-consequence (CC) and common-ratio (CR) choice pairs mimicking the original Allais example that are designed either to cross the inferred threshold partition (“Crossing” set) or to remain within a single region (“Non-Crossing” set). Appendix \ref{app:stage3-construction} reports the construction of the choice pairs and other details.

In all sessions participants first read written instructions describing the lotteries, payoffs, and the software interface. They then completed a short comprehension quiz; only those who answered all quiz questions correctly proceeded to the main tasks. The order of stages was fixed: Stage 1, then Stage 2, followed by Stage 3 and a short unincentivized questionnaire. Payoffs were incentivized using the Random Problem Selection mechanism  \autocite{azrieli2018,azrieli2020} to ensure incentive compatibility and to minimize income effects. At the end of the experiment, one task was randomly selected for payment, and the chosen option in that task was played out for real money, in addition to a fixed show-up fee.

We recruited 150 participants through Prolific in March 2024 for the main experiment session. All recruited participants were adults with a median age of 25.5 and were fluent in English. Five laboratory sessions were conducted at Chapman University in California, USA between September 2022 to March 2023. We recruited 95 student participants from the participant pool of the Economic Science Institute (ESI) Laboratory. In all sessions the experiment was conducted using the same software interface written with oTree \autocite{oTree-chen2016} with varying parameters.\footnote{We varied the number of switches allowed and the size of the stakes. The online session used points with each point converted to \$0.01 in addition to the show-up fee of \$6. The laboratory sessions displayed prizes in real dollar value in addition to the show-up fee of \$7 per the lab policy.}  The laboratory study was approved by the IRB of Chapman University. The online study was approved by the Human Ethics Office of the University of Manitoba and the IRB of Fairfield University. All participants provided informed consent.

\subsection{Main Results on Worst-Outcome-Shift}

In this section we report the main empirical regularities from sessions that implemented the one-switch protocol in Stages 1 and 2: the online Prolific session and three laboratory sessions. Table \ref{tab:summary_statistics} summarizes the overall results from Stages 1 and 2. Two earlier laboratory sessions allowed unrestricted switching and were used primarily to validate the interface and explore multiple switching; we report them separately in Appendix \ref{app:unrestricted-switching}. We also report exploratory heterogeneity by self-reported gender in the online session in Appendix \ref{app:gender-heterogeneity}; the main patterns remain qualitatively unchanged.

\begin{table}[htbp]
    \centering
    \begin{threeparttable}
    \begin{tabular}{lcc}
        \toprule
        Sample & Stage 1 switch rate & Stage 2 switch rate \\
        \midrule
        Online      & 0.520 (78/150)  & 0.613 (92/150) \\
        Laboratory    & 0.627 (42/67)   & 0.731 (49/67)  \\
        \midrule
        Pooled                 & 0.553 (120/217) & 0.650 (141/217) \\
        \bottomrule
    \end{tabular}
    \caption{Switching in Stages 1 and 2}
    \label{tab:summary_statistics}
    \begin{tablenotes}[flushleft]
    \footnotesize
    \item Notes: The table reports switching behavior in Stages 1 and 2 for sessions that implemented the one-switch protocol. “Switch rate” denotes the share of participants who choose different options across the ten-row menu.
    \end{tablenotes}
    \end{threeparttable}
\end{table}

\begin{regularity}[WS in Stage 1]
    In the online session, 78 of 150 participants (52.0\%) switch across the ten-row menu in Stage 1. In the laboratory sessions, 42 of 67 participants (62.7\%) switch. Pooling across the one-switch sessions yields 120 switches out of 217 participants (55.3\%, s.e. 3.4 pp). Switching is observed in both directions: in the pooled sample 63 participants switch from Option A to B and 57 switch from B to A.
\end{regularity}

The majority of participants switched between the options, confirming that participants respond to changes in the relative size of the worst prize even when ranks and probabilities are held fixed. The switching rate is clearly above the zero switching predicted by EU/RDU/CPT/DA.

\begin{regularity}[Probability-driven switching in Stage 2]
In the online session, 92 of 150 participants (61.3\%) switch in Stage 2; in the laboratory sessions, 49 of 67 participants (73.1\%) switch. Pooling yields 141 switches out of 217 participants (65.0\%, s.e. 3.2 pp). As in Stage 1, switching occurs in both directions: in the pooled sample 82 participants switch from Option A to B and 59 switch from B to A.
\end{regularity}

Switching is also common in Stage 2, which varies only the probability \( y \) of the worst outcome fixed at 0. The next observation shows switching behavior is strongly related across stages.

\begin{regularity}[Stable individual differences across stages]
In the pooled sample, 84.2\% of Stage-1 switchers also switch in Stage 2 (101/120). Conversely, 28.4\% of Stage-2 switchers did not switch in Stage 1 (40/141). A substantial fraction of participants do not switch in either stage: 26.3\% of the pooled sample (57/217) exhibit no switching in both Stage 1 and Stage 2. The same pattern holds separately within each environment: 88.1\% of Stage-1 switchers also switch in Stage 2 in the laboratory sessions (37/42), and 82.1\% do so online (64/78).
\end{regularity}

Stage 3 provides an additional robustness exercise as described in the experiment design. We observe reversals are systematically more frequent in the Crossing set (choice tasks with lotteries on opposite sides of the threshold line) than in the Non-Crossing set (choice tasks with lotteries on one side of the threshold line). Here we report the results from the pooled sample, but the same qualitative pattern appears within each environment and conditioning on participants who switch in both Stage 1 and Stage 2. Detailed breakdowns by environment and conditions are reported in Appendix \ref{app:additional-experimental-results}.

\begin{regularity}[Linkage to Allais-type reversals in Stage 3]
In the pooled sample, the reversal rate is 18.9\% in the Crossing-CC tasks (41/217) compared with 7.8\% in the Non-Crossing-CC tasks (17/217). The contrast is sharper for the CR tasks: 34.6\% reverse in the Crossing-CR tasks (75/217) versus 3.7\% in the Non-Crossing-CR tasks (8/217).
\end{regularity}

The raw odds ratios are sharply different across blocks: about 14 for CR and about 2.8 for CC. A pooled logit regression at the task-pair level (four observations per participant, standard errors clustered by participant) confirms that reversals are substantially more likely in the Crossing set even after conditioning on $\hat d$, $\hat\tau$, and gender, with an estimated odds ratio of about 4.1 ($p < 0.001$). Appendix \ref{app:stage3-logit} reports the full specification and estimates.

\section{Empirical Implications and Concluding Remarks}
\label{sec:remark}
 In this section we conclude with a few remarks on where the ECU framework can be especially relevant. First, we turn our attention to a long-standing empirical puzzle in insurance and small-stakes risk that behavior often looks too risk averse at small stakes for smooth EU \autocite{rabin2000}. \textcite{machina2001} emphasizes that smooth preference functionals (e.g., EU with standard differentiability) cannot exhibit payoff kinks, whereas rank-dependent models cannot avoid locally nonseparable payoff kinks. At the same time, recent experimental evidence challenges rank dependence as a general descriptive mechanism, as we argued earlier with \cite{Bernheim2020} and others. 

This combination of empirical motivation for kinked local behavior and mixed evidence for rank dependence naturally motivates studying rank-independent kink mechanisms, which is exactly where ECU framework lies. Machina notes that a generic way to generate payoff kinks is through induced preferences that take an “upper envelope” of finitely many smooth preference functionals arising from a discrete auxiliary choice. Machina's notion of kinks refers to nonsmoothness of a preference functional in payoff space, but ECU can generate the kink-like behavior through a different mechanism in the same spirit yet rank-independent. To see this, note that even a fixed prize can be evaluated using different utility functions under ECU. Thus, even though the evaluation is still that of EU and is smooth in the usual sense within a fixed context $\pi$, the kink-like behavior can still arise if the changes in the payoffs invoke switching the context.

A recent empirical finding sharpens this motivation. \textcite{dembo2025} run a three-dimensional budget-set experiment that allows a nonparametric decomposition of EU violations into failures of ordering/FOSD and failures of independence. For the vast majority of their subjects, the former dominate, limiting the explanatory gain of non-EU models that maintain FOSD by construction.\footnote{\textcite{dembo2025} define EUT-rationalizability as EU with an increasing Bernoulli index, so their decomposition separates a strengthened EU (EU plus monotonicity) rather than the raw vNM axioms. A subject with EU preferences but a non-monotone Bernoulli index would be classified as having ordering/FOSD failures rather than independence failures, indistinguishable from a genuinely non-EU subject.} The ECU is distinctively positioned here. The general ECU guarantees FOSD when Conditions 1 and 2 are imposed; without them, context-dependent utility can generate FOSD-inconsistent choice patterns through multiple utility functions. The ECU can therefore speak to both types of violation that \textcite{dembo2025} identify, unlike models that rule out FOSD failures by construction.

This connects to our earlier observation (Section~\ref{sec:axiom_interpretation}) that standard Independence fuses within-context linearity and cross-context calibration, two requirements that separate only when the context dimension is nontrivial. \textcite{dembo2025} identify an analogous dimension-dependent collapse in the revealed-preference setting: WARP and GARP are equivalent with two goods but separate with three. Similarly, when $\tilde{d} = b$, FOSD consistency and independence are entangled; introducing $\tilde{d} \neq b$ separates them, exposing behavior that mimics FOSD violations in any framework that does not model context. Our experiment's one-switch restriction suppresses the ordering violations \textcite{dembo2025} document, and investigating the ECU without Conditions 1 and 2 in richer choice environments is a natural direction for future work.

Lastly, we would like to conclude with a short discussion regarding multiple switching behavior (MSB) as a prediction target that may benefit from future research. Chew et al. interpret regular MSB as consistent with deliberate randomization induced by non-EU structure (convex preference over lotteries) rather than pure error. While \textcite{machina2001} focuses on payoff-space kinks and Chew et al. focus on mixture attraction in choice lists, both point to systematic departures from smooth EU that can generate locally sharp comparative statics. This motivates a testable conjecture: in sufficiently long choice lists, ECU-type context boundaries should generate local mixture attraction (and potentially regular MSB, though not all MSB in general) in the parameter regions where the implied context boundary intersects the list.

\pagebreak
\begin{sloppypar}
\printbibliography
\end{sloppypar}

\pagebreak

\appendix
\appendixpage
\section{Proofs}
\label{appendix:proofs}

\subsection*{Proof of Proposition 1}
    \begin{proof}
        Take $p,q\in \mathcal{L}$ such that $p$ first-order stochastically dominates $q$. Let $\pi=p([w,d])$ and $\mu=q([w,d])$. One equivalent definition of first-order stochastic dominance says that, for any non-decreasing function $u:X\rightarrow \mathbb{R}$, we have $\int u\; dp\geq \int u \; dq$. Because we have an ECU representation with Condition \ref{condition:non-decreasing-u}, $u_\mu$ is non-decreasing. Therefore, $\int u_\mu\; dq\leq \int u_\mu\; dp$. Also by definition of first-order stochastic dominance, $\pi\leq \mu$. Ergo, $u_\pi(x)\geq u_\mu(x)$ for all $x\in X$. This means that $\int u_\pi\; dp\geq \int u_\mu\; dp$. This gives us $\int u_\pi\; dp\geq \int u_\mu\; dq$, and so $p\succsim q$.
    \end{proof}

\subsection*{Proof of Lemma \ref{lemma:component-solvability}}
    \begin{proof}
        Take any $x\in X$ and $\alpha\in [0,1]$. First, consider the case where we have $y=w$. The lemma is immediately true if $\alpha=0$ because we can take $\gamma=1$. So suppose $\alpha\neq 0$. Consider $\alpha\delta_x +(1-\alpha)\delta_y=\alpha\delta_x +(1-\alpha)\delta_w$. By Axiom 4, there exists $\eta\in [0,1]$ such that $\alpha\delta_x +(1-\alpha)\delta_w\sim \eta\delta_b +(1-\eta)\delta_w$. We know from Axiom 2 that $\eta$ is a unique value. 

        We claim that $\eta\leq \alpha$. If this inequality is not true, we have $\eta>\alpha$. This implies that $\eta\delta_b +(1-\eta)\delta_w\succ \alpha\delta_b+(1-\alpha)\delta_w\succsim \alpha\delta_x+(1-\alpha)\delta_w$. The strict preference is implied by Axiom 2 while the weak preference is implied by Axiom 3. This contradicts $\alpha\delta_x +(1-\alpha)\delta_w\sim \eta\delta_b +(1-\eta)\delta_w$ and therefore establishes the claim.

       Because $\eta\leq \alpha$ and $\alpha\neq 0$, we can define $\gamma=\frac{\eta}{\alpha}\in [0,1]$. This gives us $\alpha\gamma \delta_b+\alpha(1-\gamma)\delta_w+(1-\alpha)\delta_w=\eta\delta_b+(1-\eta)\delta_w\sim \alpha\delta_x +(1-\alpha)\delta_w,$ which is what we needed to show.

        Next, take the case where $y=b.$ The lemma is immediately true if $\alpha=0$ because we can take $\gamma=1$. Suppose then that $\alpha\neq 0$. Consider $\alpha\delta_x +(1-\alpha)\delta_y=\alpha\delta_x +(1-\alpha)\delta_b$. Axiom 4 implies that there exists $\eta\in [0,1]$ such that $\alpha\delta_x +(1-\alpha)\delta_b\sim (1-\eta)\delta_b +\eta\delta_w$. We know from Axiom 2 that $\eta$ is a unique value. 

        We claim that $\eta\leq \alpha$. If this inequality is not true, we have $\eta>\alpha$. This implies that $(1-\eta)\delta_b +\eta\delta_w\prec (1-\alpha)\delta_b+\alpha\delta_w\precsim (1-\alpha)\delta_b+\alpha\delta_x$. The strict preference is implied by Axiom 2 while the weak preference relation is implied by Axiom 3. This contradicts $\alpha\delta_x +(1-\alpha)\delta_b\sim (1-\eta)\delta_b +\eta\delta_w$ and therefore establishes the claim.

        Because $\eta\leq \alpha$ and $\alpha\neq 0$, we can define $\gamma=1-\frac{\eta}{\alpha}\in [0,1]$. This gives us $\alpha\gamma \delta_b+\alpha(1-\gamma)\delta_w+(1-\alpha)\delta_b=(1-\eta)\delta_b+\eta\delta_w\sim \alpha\delta_x +(1-\alpha)\delta_b$, which is what we needed to show.
    \end{proof}

\subsection*{More Lemmas}
We next provide additional lemmas that are useful for proving Theorem 1. The first lemma shows that the parameter we recover using set $\mathcal{D}$ comes from the ECU representation. 
    \begin{lemma}
        If $\succsim$ has an ECU representation, then $\tilde{d}=d$.
    \end{lemma} 
    \begin{proof}
        Take the case where $d\neq b$. We show that $x\in {\mathcal{D}}\iff x\in[w,d]$. First, to show that $x\in [w,d]\implies x \in\mathcal{D}$, take $x\in [w,d]$. Because $\delta_x([w,d])=1$, $V(\delta_x)=u_1(x)$. Therefore, $\phi_x$ is the value such that $u_1(x)=\phi_xu(b)+(1-\phi_x)u(w)$. To see this, if we let $p=\phi_x\delta_b+(1-\phi_x)\delta_w$, then $V(p)=\phi_xu(b) +(1-\phi_x)u(w)=u_1(x)=V(\delta_x),$ so that $\delta_x\sim \phi_x\delta_b+(1-\phi_x)\delta_w$. Now, take any $\alpha\in[0,1]$. and define lotteries $q=\alpha\delta_x+(1-\alpha)\delta_w $ and $r=\alpha \phi_x \delta_b+\alpha(1-\phi_x)\delta_w+(1-\alpha)\delta_w$. Because $q$ places probability 1 on disappointing prizes,
        \begin{align*}
            V(q)&=\alpha u_1(x)+(1-\alpha)u_1(w) \\
            &=\alpha\phi_x u(b)+\alpha(1-\phi_x)u(w)+(1-\alpha)u(w)\\
            &=V(\alpha \phi_x \delta_b+\alpha(1-\phi_x)\delta_w+(1-\alpha)\delta_w). \\
        \end{align*}
        This gives us $q\sim r$, and so $x\in \mathcal{D}$.

        For the other direction, take any  $x\in (d,b)$. Therefore, $\phi_x$ is the value such that $u_0(x)=\phi_xu(b)+(1-\phi_x)u(w)$. To see this, if we let $p'=\phi_x\delta_b+(1-\phi_x)\delta_w$, then $V(p')=\phi_xu(b) +(1-\phi_x)u(w)=u_0(x)=V(\delta_x).$

        By assumption of an ECU representation, there exists a $\beta\in (0,1)$ such that $u_\beta(x)\neq u_0(x)$. If we let  $q'=(1-\beta)\delta_x+\beta\delta_w$, we have 
        \begin{align*} 
            V(q')&=(1-\beta)u_\beta(x)+\beta u_\beta(w)\\
            &\neq (1-\beta)u_0(x)+\beta u_0(w)\\
            &=(1-\beta) \phi_x u(b)+(1-\beta)(1-\phi_x)u(w)+\beta u(w)\\
            &=V((1-\beta)\phi_x\delta_b+(1-\beta)(1-\phi_x)\delta_w+\beta  \delta_w).
        \end{align*}
        This means that we do not have $(1-\beta)\delta_x+\beta\delta_w\sim (1-\beta)\phi_x\delta_b+(1-\beta)(1-\phi_x)\delta_w+\beta  \delta_w$, and so $x\notin \mathcal{D}$. This establishes that ${\mathcal{D}}=[w,d]$, which implies that  $\sup{\mathcal{D}}=\tilde{d}=d$.

        Next, consider the case where $d=b$. Take any $x\in [w,b)$. The same argument for the case with $d\neq b$ and $x\in [w,d]$ shows that $x\in \mathcal{D}$. This gives us ${\mathcal{D}}=[w,b)$, and so $\sup{\mathcal{D}}=\tilde{d}=b=d$.
    \end{proof}

    \begin{lemma}
        Suppose $\succsim$ has an ECU representation. If $\tilde{d}\neq b$ then, for any $x\in [w,d]$ and $\alpha\in (0,1]$, we have $u_\alpha(x)=\phi_x^\alpha u(b)+(1-\phi_x^\alpha)u(w)$.
    \end{lemma}
    \begin{proof}
        Take any $x\in[w,d]$ and $\alpha\in(0,1]$. By Lemma 2, $\tilde{d}=d$, and so $\phi_x^\alpha$ is the value such that $\alpha\delta_x+(1-\alpha)\delta_b\sim \alpha\phi_x^\alpha\delta_b+\alpha(1-\phi_\alpha^x)\delta_w+(1-\alpha)\delta_b$. This means that
        \begin{align*}
            \alpha u_\alpha(x)+(1-\alpha)u(b)&=V(\alpha\delta_x+(1-\alpha)\delta_b) \\
            &=V(\alpha\phi_x^\alpha\delta_b+\alpha(1-\phi^\alpha_x)\delta_w+(1-\alpha)\delta_b)   \\  
            &=\alpha\phi_x^\alpha u(b)+\alpha(1-\phi_x^\alpha)u(w)+(1-\alpha)u(b).\\
        \end{align*}
        Applying algebra to the equation gives us $u_\alpha(x)=\phi_x^\alpha u(b)+(1-\phi_x^\alpha)u(w).$
    \end{proof}

    \begin{lemma}
        Suppose that $\succsim$ has an ECU representation. If $\tilde{d}\neq b$ then, for any $x\notin [w,d]$ and $\alpha\in [0,1)$, we have $u_\alpha(x)=\phi_x^\alpha u(b)+(1-\phi_x^\alpha)u(w)$.
    \end{lemma}
    \begin{proof}
        Take any $x\notin[w,d]$ and $\alpha\in[0,1)$. By Lemma 2, $\tilde{d}=d$, and so $\phi_x^\alpha$ is the value such that $\alpha\delta_w+(1-\alpha)\delta_x\sim\alpha\delta_w+(1-\alpha)\phi_x^\alpha\delta_b+(1-\alpha)(1-\phi_x^\alpha)\delta_w$. This means that
        \begin{align*}
            \alpha u(w)+(1-\alpha)u_\alpha(x)&=V(\alpha\delta_w+(1-\alpha)\delta_x) \\
            &=V(\alpha\delta_w+(1-\alpha)\phi_x^\alpha\delta_b+(1-\alpha)(1-\phi_x^\alpha)\delta_w)   \\  
            &=\alpha u(w)+(1-\alpha)\phi_x^\alpha u(b)+(1-\alpha)(1-\phi_x^\alpha)u(w) .\\
        \end{align*}
        Applying algebra to the equation gives us $u_\alpha(x)=\phi_x^\alpha u(b)+(1-\phi_x^\alpha)u(w).$
    \end{proof}

    \begin{lemma}
        Suppose $\succsim$ has an ECU representation.	If $\tilde{d}=b$ then, for any $x\in [w,b]$, we have $u_1(x)=\phi_x^1 u_1(b)+(1-\phi_x^1)u_1(w)$.
    \end{lemma}
    \begin{proof}
        Take any $x\in[w,b]$. Because $\tilde{d}=b$, $\phi_x^1$ is the value such that $\delta_x\sim \phi_x^1\delta_b+(1-\phi_x^1)\delta_w$. We also know by Lemma 2 that $b=\tilde{d}=d$. Ergo, $u_1(x)=V(\delta_x)=V(\phi_x^1\delta_b+(1-\phi^1_x)\delta_w) =\phi_x^1 u_{1}(b)+(1-\phi_x^1)u_{1}(w).$   
    \end{proof}

\subsection*{Proof of Theorem 1}
    \begin{proof}	
    \textbf{Sufficiency:} 

    \noindent 
    We first prove the sufficiency of the axioms for an ECU representation. Suppose that Axioms 1-5 and the regularity condition are satisfied. 	

    \noindent 
    \textbf{Case 1: $\tilde{d}\neq b$}

    Let $\tilde{d}=d$ for the threshold of the ECU representation. By definition of $\tilde{d}$, it must be in $[w,b)$. We let $u_\pi(w)=0$ for all $\pi\in(0,1]$. Also let $u_\pi(b)=1$ for all $\pi\in[0,1)$. For any $x\in (w,d]$ and $\pi\in(0,1]$, let $u_\pi(x)=\phi_x^\pi$. For any $x\in (d,b)$ and $\pi\in[0,1)$, let $u_\pi(x)=\phi_x^\pi$. 

    The requirement that $u_\pi(w)<u_\pi(b)$ for all $\pi \in[0,1]$ such that $u_\pi(w), u_\pi(b)$ are defined is clearly met. It is also easy to see that, for any $x\in\{w,b\}$ and $\pi,\mu\in[0,1]$ such that $u_{\pi}(x), u_{\mu}(x)$ are defined, $u_\pi(x)=u_\mu(x) $. Because $\phi_x^\pi\in [0,1]$ for all $x$ and $\pi$ such that $\phi_x^\pi$ are defined, $u_\pi(x)\in [u(w),u(b)]$ for all $x$ and $\pi$ such that $\phi_x^\pi$ are defined. 

    For any $x\in(w,b)$, because we assumed that there exists $\pi,\mu\in(0,1)$ such that $\phi_x^\pi\neq \phi_x^\mu$, there exist $\pi,\mu\in(0,1)$ such that $u_\pi(x)\neq u_\mu(x)$. Therefore, we have a set of contextual utilities.

    We want to show that, for any $p,q\in \mathcal{L}$, we have $p\succsim q \iff V(p)\geq V(q)$. Suppose that $p([w,d])=\alpha$ and $q([w,d])=\gamma$. We have
    \begin{align*}
        p\succsim q &\iff \sum_{x\in supp(p)} p(x)\left[\phi_x^\alpha\delta_b+(1-\phi^\alpha_x)\delta_w\right]\succsim \sum_{x\in supp(q)} q(x)\left[\phi^\gamma_x\delta_b+(1-\phi^\gamma_x)\delta_w\right] \text{ by Axiom 5}\\
        &\iff \sum_{x\in supp(p)} p(x)\phi^\alpha_x\geq \sum_{x\in supp(q)} q(x)\phi^\gamma_x\text{ by Axiom 2}\\
        &\iff \sum_{x\in supp(p)} p(x)u_\alpha(x)\geq \sum_{x\in supp(q)} q(x)u_\gamma(x)\\
        &\iff \int u_\alpha\; dp \geq \int u_\gamma \; dq.\\	   	  
    \end{align*}

    \noindent
    \textbf{Case 2: $\tilde{d}=b$} 

    Let $\tilde{d}=b$ for the threshold of the ECU representation. We let $u_\pi(w)=0$ and $u_\pi(b)=1$ for all $\pi\in(0,1]$. For any $x\in (w,b)$ and $\alpha\in (0,1]$, we let $u_\alpha(x)=\phi_x^1$. 

    As in Case 1, the requirement that $u_\pi(w)< u_\pi(b)$ for all $\pi \in[0,1]$ such that $u_\pi(w), u_\pi(b)$ are defined is clearly met. It is also easy to see that for any $x\in\{w,b\}$ and $\pi,\mu\in[0,1]$ such that $u_{\pi}(x), u_{\mu}(x)$ are defined, $u_\pi(x)=u_\mu(x) $. Because $\phi_x^\pi\in [0,1]$ for all $x$ and $\pi$ such that $\phi_x^\pi$ are defined, $u_\pi(x)\in [u(w),u(b)]$ for all $x$ and $\pi$ such that $u_\pi(x)$ are defined. This gives us a set of utility functions that is contextual.

    The argument for $p\succsim q\iff V(p)\geq V(q)$ is identical to that of the first case. This completes proof of sufficiency for Case 2.

    \textbf{Necessity:}

    We move on to prove necessity. Suppose that preference relation $\succsim$ has an ECU representation. The necessity of Axiom 1 is trivial so the proof is omitted.

    For Axiom 2, take any $0\leq t_2\leq t_1\leq1$. We have
    \begin{align*}
        t_1\delta_b+(1-t_1)\delta_w\succsim t_2\delta_b+(1-t_2)\delta_w&\iff t_1u(b)+(1-t_1)u(w)\geq t_2u(b)+(1-t_2)u(w)\\
        & \iff t_1\geq t_2\\  
    \end{align*}
    The second line is implied by $u(b)> u(w)$.

    We next prove the necessity of Axiom 3. Take any $x\in X$ and $\alpha\in [0,1]$. We have $V(\alpha\delta_b+(1-\alpha)\delta_w)=\alpha u(b)+(1-\alpha)u(w)$. Let $p=\alpha\delta_b+(1-\alpha)\delta_x$. Let $\pi=p([w,d])$. Then $V(p)=\alpha u_\pi(b)+(1-\alpha)u_\pi(x)=\alpha u(b)+(1-\alpha)u_\pi(x)$. Because $u_\pi(x)\in[u(w),u(b)]$,
    we have $V(p)=\alpha u(b)+(1-\alpha)u_\pi(x)\geq \alpha u(b)+(1-\alpha)u(w)$. This gives us $\alpha\delta_b+(1-\alpha)\delta_x\succsim \alpha\delta_b+(1-\alpha)\delta_w$.

    Next, let $q=\alpha\delta_x+(1-\alpha)\delta_w$. Let $\nu=q([w,d])$. Then $V(q)=\alpha u_\nu(x)+(1-\alpha)u_\nu(w)=\alpha u_\nu(x)+(1-\alpha)u(w)$. Because $u_\nu(x)\in[u(w),u(b)]$,
    we have $V(q)=\alpha u_\nu(x)+(1-\alpha)u(w)\leq \alpha u(b)+(1-\alpha)u(w)$. This gives us $ \alpha\delta_b+(1-\alpha)\delta_w \succsim  \alpha\delta_x+(1-\alpha)\delta_w$. We conclude that Axiom 3 is satisfied.

    For the necessity of Axiom 4, take any $p\in \mathcal{L}$. Let $p=(x_1,p_1;\ldots;x_n,p_n)$ and  $p([w,d])=\rho$. Then $$V(p)=\sum_{i=1}^n p_iu_\rho(x_i).$$ Because $u_\rho(x)\in [u(w),u(b)]$ for all $x\in X$, $V(p)\in [u(w),u(b)]$. An application of the Intermediate Value Theorem thus tells us that there exists $\gamma$ such that $V(p)=\gamma u(b)+(1-\gamma)u(w)$. Let $q=\gamma\delta_b+(1-\gamma)\delta_w$. Then $V(q)=\gamma u(b)+(1-\gamma)u(w)=V(p)$ so that $p\sim q$.

    The necessity of Axiom 5 follows. We prove this by cases. Take the case where $\alpha=0$ and let $p$ be such that $p([w,\tilde{d}])=\alpha=0$. This must mean that $\tilde{d}\neq b$. By Lemma 2, this also means that $p([w,d])=0$. If $d=b$ for the ECU representation, then $q([w,d])=1$ for all $q\in \mathcal{L}$. Because $p([w,d])=0\neq 1$, we know $d\neq b$. It must also be the case that $x\notin[w,d]$ for all $x\in supp(p)$. We thus have 
    \begin{align*}
        V(p)&=\sum_{x\in supp(p)}p(x)u_0(x)         \\
        &=  \sum_{x\in supp(p)}p(x)\left[\phi_x^0u(b)+(1-\phi_x^0)u(w)\right]         \\
        &=V(\sum_{x\in supp(p)}p(x)\left[\phi_x^0\delta_b+(1-\phi_x^0)\delta_w\right] )   \\
    \end{align*}
    $\implies p\sim \sum_{x\in supp(p)}p(x)\left[\phi_x^0\delta_b+(1-\phi_x^0)\delta_w\right]$.
    The second line is implied by Lemma 4. 

    Next, take the case where $\alpha=1$ and let $p$ be such that $p([w,\tilde{d}])=\alpha=1$. By Lemma 2, this means that $p([w,d])=1$. This must in turn mean that $x\in[w,d]$ for all $x\in supp(p)$. We thus have 
    \begin{align*}
        V(p)&=\sum_{x\in supp(p)}p(x)u_1(x)         \\
        &=  \sum_{x\in supp(p)}p(x)\left[\phi_x^1u(b)+(1-\phi_x^1)u(w)\right]         \\
        &=V(\sum_{x\in supp(p)}p(x)\left[\phi_x^1\delta_b+(1-\phi_x^1)\delta_w\right] )   \\
    \end{align*}
    $\implies p\sim \sum_{x\in supp(p)}p(x)\left[\phi_x^0\delta_b+(1-\phi_x^0)\delta_w\right]$.
    The second line is implied by Lemma 3 if $\tilde{d}\neq b$ or by Lemma 5 if $\tilde{d}=b$. 

    For the last case, take $\alpha\in (0,1)$ and let $p$ be such that $p([w,\tilde{d}])=\alpha$. Because $\alpha\neq 1$, $\tilde{d}\neq b$. By Lemma 2, we also have $p([w,d])=\alpha$. If $d=b$ for the ECU representation, then $q([w,d])=1$ for all $q\in \mathcal{L}$. Because $p([w,d])\neq 1$, we know that $d\neq b$. We then have 
    \begin{align*}
        V(p)&=\sum_{x\in supp(p)}p(x)u_\alpha(x)         \\
        &=  \sum_{x\in supp(p)}p(x)\left[\phi_x^\alpha u(b)+(1-\phi_x^\alpha)u(w)\right]         \\
        &=V(\sum_{x\in supp(p)}p(x)\left[\phi_x^\alpha\delta_b+(1-\phi_x^\alpha)\delta_w\right] )   \\
    \end{align*}
    $\implies p\sim \sum_{x\in supp(p)}p(x)\left[\phi_x^0\delta_b+(1-\phi_x^0)\delta_w\right]$.
    The second line is implied by a combination of Lemma 3 and Lemma 4. 

    Finally, we prove the necessity of the regularity condition. Suppose that $\tilde{d}\neq b$. Because $\tilde{d}\neq b$, Lemma 2 tells us that $d\neq b$. Consider the case where $x\in(w,d]$. Because the set of utility functions is contextual, there exist $\pi,\mu\in(0,1)$ such that $u_\pi(x)\neq u_\mu(x)$. In combination with Lemma 3, this means that
    \begin{align*}
        &\phi_x^\pi u_\pi(b)+(1-\phi_x^\pi)u_\pi(w)\neq \phi_x^\mu u_\mu(b)+(1-\phi_x^\mu)u_\mu(w) \\
        \implies& \phi_x^\pi u_\pi(b)+(1-\phi_x^\pi)u_\pi(w)\neq \phi_x^\mu u_\pi(b)+(1-\phi_x^\mu)u_\pi(w) \\
        \implies&\phi_x^\pi\neq \phi_x^\mu\\
    \end{align*}
    For the case where $x\in (d,b)$, we can use an identical argument using Lemma 4 instead of Lemma 3.   
    \end{proof} 

\subsection*{Proof of Proposition 2}
\begin{proof}
	Take any $p,q,r\in \cal{L}$ such that $p([w,d])=q([w,d])=r([w,d])$ and any $\gamma\in [0,1]$. Suppose $p\succsim q$. Let $\alpha$ be the common value for $p([w,d])=q([w,d])=r([w,d])$. Since $\succsim$ has an ECU representation, we know that $\int u_\alpha \text{ }dp\geq \int u_\alpha \text{ }dq$. 
	
	Now, let $p'=\gamma p+(1-\gamma)r$ and $q'=\gamma q+(1-\gamma)r$. We know that $p'([w,d])=\gamma \alpha +(1-\gamma )\alpha=\alpha$. We also know that $q'([w,d])=\alpha$. Therefore, we have \begin{align*}
	V(p')=\int u_\alpha \text{ }dp'&=\sum_{x\in supp(p')}u_\alpha(x)p'(x)\\&=\left[\alpha \sum_{x\in supp(p)}u_\alpha(x)p(x)\right]+\left[(1-\alpha) \sum_{x\in supp(r)}u_\alpha(x)r(x)\right],\\
	\end{align*} and similarly, $$V(q')=\int u_\alpha \text{ }dq'=\left[\alpha \sum_{x\in supp(q)}u_\alpha(x)q(x)\right]+\left[(1-\alpha) \sum_{x\in supp(r)}u_\alpha(x)r(x)\right].$$ This tells us that we have $$V(p')\geq V(q')\iff \sum_{x\in supp(p)}u_\alpha(x)p(x)\geq \sum_{x\in supp(q)}u_\alpha(x)q(x)\iff \int u_\alpha dp\geq \int u_\alpha dq.$$ Since the last inequality is true, we conclude that $V(p')\geq V(q')$. Since we have an ECU representation, this means that $p'=\gamma p+(1-\gamma)r\succsim \gamma q+(1-\gamma)r=q'$.
\end{proof}

\subsection*{Proof of Theorem 2}
    \begin{proof}	
    The "if" direction is immediate by the linearity of integrals. So, we need only to prove the "only if" direction. Suppose that a preference relation $\succsim$ has an ECU with $\lbrace d_1,(u_\alpha)_{\alpha\in [0,1]} \rbrace$ and $\lbrace d_2,(v_\alpha)_{\alpha\in [0,1]} \rbrace$ . For value $V(p)$  of any lottery $p$ given in the definition of an ECU, we use the notation $V_1(p)$ when using the ECU with parameters $\lbrace d_i,(u_\alpha)_{\alpha\in [0,1]}\rbrace$, while we use $V_2(p)$ when we use the ECU with parameters $\lbrace d_2,(v_\alpha)_{\alpha\in [0,1]}\rbrace$. As was done with functions $u_\alpha$, we often simplify and write $v(w)$ or $v(b)$ for the value of $v_\alpha(w)$ and $v_\alpha(b)$, which are common across all $\alpha\in [0,1]$.

    Suppose that $d_1\neq d_2$. WLOG assume $d_1<d_2$. Take any $x\in (d_1,d_2)$. Let $\phi_x\in [0,1]$ be the value such that $v_1(x)=\phi_xv(b)+(1-\phi_x)v(w)$, meaning that $\delta_x\sim \phi_x\delta_b+(1-\phi_x)\delta_w$. Such a value exists by an application of the Intermediate Value Theorem. For any $\alpha \in [0,1]$ we have 
    \begin{align*}
        V_2(\alpha\delta_x+(1-\alpha)\delta_w )&= \alpha v_1(x)+(1-\alpha)v_1(w)    \\
        &=\alpha \phi_xv(b)+\alpha (1-\phi_x)v(w)+(1-\alpha)v(w) \\
        &=V_2\left(\alpha \phi_x \delta_b+\alpha(1-\phi_x)\delta_w+(1-\alpha)\delta_w\right),
    \end{align*}
    so that $\alpha\delta_x+(1-\alpha)\delta_w\sim \alpha \phi_x \delta_b+\alpha(1-\phi_x)\delta_w+(1-\alpha)\delta_w.$

    Inamuch as we have $\delta_x\sim \phi_x\delta_b+(1-\phi_x)\delta_w$, it must also be the case that $u_0(x)=\phi_xu(b)+(1-\phi_x)u(w)$. Because we have an ECU with $d_1\neq b$, there exists $\alpha\in (0,1)$ such that $u_\alpha(x)\neq u_0(x)$. This tells us that
    \begin{align*}
        V_1((1-\alpha)\delta_x+\alpha\delta_w )&= (1-\alpha)u_\alpha(x)+\alpha u(w)    \\
        &\neq (1-\alpha)u_0(x)+\alpha u(w)   \\
        &=  (1-\alpha)\phi_x u(b)+(1-\alpha)(1-\phi_x) u(w)+\alpha u(w)        \\
        &=V_1\left((1-\alpha) \phi_x \delta_b+(1-\alpha)(1-\phi_x)\delta_w+\alpha\delta_w\right),
    \end{align*}
    so that we do not have $(1-\alpha)\delta_x+\alpha\delta_w\sim (1-\alpha) \phi_x \delta_b+(1-\alpha)(1-\phi_x)\delta_w+\alpha\delta_w$. This contradicts $\lbrace d_1,(u_\alpha)_{\alpha\in [0,1]} \rbrace$ and $\lbrace d_2,(v_\alpha)_{\alpha\in [0,1]} \rbrace$ representing the same preferences. We conclude that $d_1=d_2$. For the remainder of the proof, we refer to the common value as $d$.

    We next show that $v_\alpha=ku_\alpha+c$ for all $\alpha\in [0,1]$ with $k\in \mathbb{R}_{++}$ and $c\in \mathbb{R}$. We prove this in two steps. First, we show that, for any $\alpha \in (0,1]$ and $x\in [w,d]$, we have $v_\alpha(x)=ku_\alpha(x)+c$. In the second step, we show that for any $\alpha \in [0,1)$ and $x\in [w,d]$, we have $v_\alpha(x)=ku_\alpha(x)+c$.In the first step, we do not need $\alpha=0$ because $u_0$ and $v_0$ are not defined on $[w,d]$. Similarly, for the second step, we do not consider $\alpha=1$ because $u_1,v_1$ are not defined on $X\setminus [w,d]$. 

    Take any $\alpha \in (0,1]$ and $x\in [w,d]$. Let $p=(x,\alpha; b,1-\alpha)$. We have $V_1(p)=\alpha u_\alpha(x)+(1-\alpha) u(b).$ Letting $\gamma_x= \frac{\alpha\left[ u_\alpha(x)-u(b)\right]}{u(w)-u(b)}$, we have $\gamma_xu(w)+(1-\gamma_x)u(b)=V_1(p)$. It is easy to see that $\gamma_x\in [0,1]$. Therefore, letting $q=(w,\gamma_x;b,1-\gamma_x)$, we have $p\sim q$. Because the two ECU representations represent the same preferences, we must also have 
    \begin{align*}
        &\alpha v_\alpha(x)+(1-\alpha)v(b)=\gamma_xv(w)+(1-\gamma_x)v(b)\\
        \implies &v_\alpha(x)=u_\alpha(x)\left[\frac{v(w)-v(b)}{u(w)-u(b)} \right]+  \frac{v(b)u(w)-u(b)v(w)}{u(w)-u(b)} . \\
    \end{align*}
    We let $k=\left[\frac{v(w)-v(b)}{u(w)-u(b)} \right]$ and $c=\frac{v(b)u(w)-u(b)v(w)}{u(w)-u(b)}  $. It is easy to see that $k\in \mathbb{R}_{++}.$

    Now, take any $\alpha \in [0,1)$ and $x\notin [w,d]$. Let $p=(w,\alpha; x,1-\alpha)$. We have $V_1(p)=\alpha u(w)+(1-\alpha) u_\alpha(x).$ Letting $\gamma_x= \frac{\alpha u(w)+(1-\alpha)u_\alpha(x)-u(b)}{u(w)-u(b)}$, we have $\gamma_xu(w)+(1-\gamma_x)u(b)=V_1(p)$. It is easy to see that $\gamma_x\in [0,1]$. Therefore, letting $q=(w,\gamma_x;b,1-\gamma_x)$, we have $p\sim q$. Because the two ECU representations represent the same preferences, we must also have 
    \begin{align*}
        &\alpha v(w)+(1-\alpha)v_\alpha(x)=\gamma_xv(w)+(1-\gamma_x)v(b)\\
        \implies &v_\alpha(x)=u_\alpha(x)\left[\frac{v(w)-v(b)}{u(w)-u(b)} \right]+  \frac{v(b)u(w)-u(b)v(w)}{u(w)-u(b)} \\
        \implies&v_\alpha(x)=ku_\alpha(x)+c.\\
    \end{align*}
    \end{proof} 

 \subsection*{Proof of Corollary 1}
\begin{proof}
	 Take any $\pi,\mu \in [0,1]$. By Theorem 2, we know that $v_\pi=ku_\pi +c$ and $v_\mu=ku_\mu +c$ for some $k\in \mathbb{R}_{++}$ and $c\in \mathbb{R}$. We have $\pi \overset{u}{\sim}\mu \iff u_\pi=u_\mu \iff ku_\pi +c=ku_\mu+c\iff v_\pi=v_\mu\iff \pi \overset{v}{\sim}\mu$. We conclude that $\overset{u}{\sim}=\overset{v}{\sim}$.
\end{proof}

\pagebreak

\section{ECU and connected CR--CC--MX patterns}\label{app:crccmx-ecu}

This appendix shows how a binary ECU can reproduce the prominent sign pattern in connected
common-ratio (CR), common-consequence (CC), and mixture (MX) problems documented by
\textcite{mcgranaghan2025}. The goal is compatibility and mechanism rather than a full
structural fit exercise.

\subsection{Connected CR--CC--MX problems and valuation objects}\label{app:crccmx-ecu:setup}

Fix prizes $H>M>0$ and parameters $(p,r)\in(0,1)^2$. Consider the connected lotteries
\[
A=(M),\qquad B=(H,p;0,1-p),\qquad
B_1=(1-r)A+rB=(H,pr;\ M-r;\ 0,r(1-p)),
\]
\[
C=(M,r;0,1-r),\qquad D=(H,pr;0,1-pr).
\]
The CR comparison is $(A,B)$ versus $(C,D)$, the CC comparison is $(A,B_1)$ versus $(C,D)$,
and the MX comparison is $(A,B)$ versus $(A,B_1)$.

Following \textcite{mcgranaghan2025}, for any monotone preference there exist indifference
cutoffs $h^*_{AB}$, $h^*_{AB_1}$, and $h^*_{CD}$ such that, holding $(M,p,r)$ fixed,
\[
A \succ B \iff H<h^*_{AB},\qquad
A \succ B_1 \iff H<h^*_{AB_1},\qquad
C \succ D \iff H<h^*_{CD}.
\]
Define the connected ``reactions'' (valuations)
\[
\Delta^*_{CR}=h^*_{AB}-h^*_{CD},\qquad
\Delta^*_{CC}=h^*_{AB_1}-h^*_{CD},\qquad
\Delta^*_{MX}=h^*_{AB}-h^*_{AB_1}.
\]
CRP corresponds to $\Delta^*_{CR}>0$, RCCP to $\Delta^*_{CC}<0$, and MXP to $\Delta^*_{MX}>0$.

\subsection{Binary ECU specialization and $\pi$-contexts}\label{app:crccmx-ecu:ecu}

Consider a binary ECU with disappointment threshold $d$ and probability threshold $\tau\in(0,1)$.
For a lottery $\ell$, define its context by
\[
\pi(\ell)=\Pr_{\ell}(x\le d).
\]
Binary ECU evaluates $\ell$ by
\[
V(\ell)=
\begin{cases}
\mathbb{E}_{\ell}[u(x)] & \text{if }\pi(\ell)\le\tau,\\
\mathbb{E}_{\ell}[v(x)] & \text{if }\pi(\ell)>\tau,
\end{cases}
\]
where $u$ and $v$ are strictly increasing (there is no rank dependence or probability weighting).

For the connected CR--CC--MX lotteries above, impose the simple condition $0<d<M$.
Then $\{x\le d\}$ coincides with $\{x=0\}$ in these lotteries, so the $\pi$-context assignment
depends only on the probability of $0$:
\[
\pi(A)=0,\qquad
\pi(B)=1-p,\qquad
\pi(B_1)=r(1-p),\qquad
\pi(C)=1-r,\qquad
\pi(D)=1-pr.
\]
Two monotonicity relations are immediate:
\[
\pi(B_1)\le \pi(B),\qquad \pi(C)\le \pi(D),
\]
so only a subset of context configurations is feasible. For convenience in what follows, we set
$u(0)=v(0)=0$.

\subsection{Region partition and sign predictions}\label{app:crccmx-ecu:partition}

Fix $\tau\in(0,1)$. The four threshold comparisons that govern $\pi$-context assignment are
\begin{equation}\label{eq:appendix-CRCCMX-four_threshold_inequalities-tight}
1-p\ \lesseqgtr\ \tau,\qquad r(1-p)\ \lesseqgtr\ \tau,\qquad 1-r\ \lesseqgtr\ \tau,\qquad 1-pr\ \lesseqgtr\ \tau,
\end{equation}
where $\lesseqgtr$ indicates either $\le$ or $>$. Because $\pi(B_1)\le\pi(B)$ and $\pi(C)\le\pi(D)$,
only the seven context configurations listed in Table~\ref{tab:crccmx-ecu-regions} are feasible.

To state these regions compactly, define the boundary curves
\begin{equation}\label{eq:appendix-CRCCMX-key_boundaries-tight}
p_0:=1-\tau,\qquad r_0:=1-\tau,\qquad
r_{B_1}(p):=\frac{\tau}{1-p}\ \ (p<p_0),\qquad
r_D(p):=\frac{1-\tau}{p}\ \ (p\ge p_0).
\end{equation}

\paragraph{Cutoffs implied by each context configuration.}
With $u(0)=v(0)=0$, define the benchmark cutoffs (functions of $(M,p)$)
\[
h_u:=u^{-1}\!\left(\frac{u(M)}{p}\right),\qquad
h_v:=v^{-1}\!\left(\frac{v(M)}{p}\right),\qquad
h_{uv}:=v^{-1}\!\left(\frac{u(M)}{p}\right).
\]
Then, in the connected problem above:
\begin{itemize}
\item $h^*_{AB}=h_u$ if $B$ is in the low-$\pi$ context, and $h^*_{AB}=h_{uv}$ if $B$ is in the high-$\pi$ context.
\item $h^*_{CD}=h_u$ if both $C$ and $D$ are in the low-$\pi$ context; $h^*_{CD}=h_{uv}$ if $C$ is low-$\pi$ and $D$ is high-$\pi$; and $h^*_{CD}=h_v$ if both are high-$\pi$.
\item $h^*_{AB_1}=h_u$ if $B_1$ is in the low-$\pi$ context (the dependence on $r$ cancels because $B_1=(1-r)A+rB$). If $B_1$ is in the high-$\pi$ context, then $h^*_{AB_1}$ exceeds $h_{uv}$ whenever $u(M)>v(M)$.
\end{itemize}

\paragraph{Maintained ordering.}
To translate context configurations into sign predictions, we maintain the transparent ordering
\[
h_u<h_v<h_{uv},
\]
which holds, for example, when $u$ is less risk-averse than $v$ (so $h_u<h_v$) and $u(M)>v(M)$
(so $h_{uv}>h_v$). Under this ordering, Table~\ref{tab:crccmx-ecu-regions} reports the signs of
$(\Delta^*_{CR},\Delta^*_{CC},\Delta^*_{MX})$ by region. In particular, the CRP--RCCP--MXP pattern
arises in Region~IV.

\begin{table}[!htbp]\centering
\begin{tabular}{p{0.26\textwidth} p{0.42\textwidth} p{0.26\textwidth}}
\hline
\textbf{Region (conditions)} &
\textbf{$\pi$-contexts induced by \eqref{eq:appendix-CRCCMX-four_threshold_inequalities-tight}} &
\textbf{Signs} \newline
\textbf{for $(\Delta^*_{CR},\Delta^*_{CC},\Delta^*_{MX})$}
\\
\hline
\multicolumn{3}{l}{\textbf{Case 1: $p\ge p_0$ (so $B$ and $B_1$ are both low-$\pi$)}}\\[2pt]
I: $r\ge r_D(p)$ &
$B,B_1,C,D$ low-$\pi$ &
$(0,0,0)$
\\
II: $r<r_0$ &
$B,B_1$ low-$\pi$;\ $C,D$ high-$\pi$ &
$(-,-)$
\\
III: $r_0\le r<r_D(p)$ &
$B,B_1,C$ low-$\pi$;\ $D$ high-$\pi$ &
$(-,-)$
\\
\hline
\multicolumn{3}{l}{\textbf{Case 2: $p<p_0$ (so $B$ is high-$\pi$; $B_1$ depends on $r$)}}\\[2pt]
IV: $r\le r_{B_1}(p)$ and $r<r_0$ &
$B$ high-$\pi$;\ $B_1$ low-$\pi$;\ $C,D$ high-$\pi$ &
$(+,-,+)$
\\
V: $r\le r_{B_1}(p)$ and $r\ge r_0$ &
$B$ high-$\pi$;\ $B_1$ low-$\pi$;\ $C$ low-$\pi$;\ $D$ high-$\pi$ &
$(0,-,+)$
\\
VI: $r>r_{B_1}(p)$ and $r<r_0$ &
$B,B_1,C,D$ high-$\pi$ &
$(+,+,-)$
\\
VII: $r>r_{B_1}(p)$ and $r\ge r_0$ &
$B,B_1$ high-$\pi$;\ $C$ low-$\pi$;\ $D$ high-$\pi$ &
$(0,+,-)$
\\
\hline
\end{tabular}
\caption{Feasible regions (aligned with the $p\ge p_0$ vs.\ $p<p_0$ split) and ECU sign predictions}
\label{tab:crccmx-ecu-regions}
\end{table}

\subsection{Worked numerical example (Region IV)}\label{app:crccmx-ecu:example}

We provide a numerical example that lies in Region~IV and yields the CRP--RCCP--MXP pattern. Let $M=10$, pick any $d$ with $0<d<M$ (e.g., $d=1$), and set $\tau=0.30$. Let
\[
u(x)=x,\qquad v(x)=\sqrt{x},
\]
so $u(0)=v(0)=0$. Choose $(p,r)=(0.20,0.30)$. Then
\begin{align*}
    \pi(B)=1-p=0.80>\tau,\qquad & \pi(B_1)=r(1-p)=0.24\le\tau, \\
    \pi(C)=1-r=0.70>\tau,\qquad & \pi(D)=1-pr=0.94>\tau,
\end{align*}

so the configuration is Region~IV.

\paragraph{Cutoffs and signs.}
Compute the benchmark cutoffs:
\[
h_u=\frac{u(M)}{p}=\frac{10}{0.2}=50,\qquad
h_v=\left(\frac{v(M)}{p}\right)^2=\left(\frac{\sqrt{10}}{0.2}\right)^2=250,\qquad
h_{uv}=\left(\frac{u(M)}{p}\right)^2=\left(\frac{10}{0.2}\right)^2=2500.
\]
Under Region~IV, $h^*_{AB}=h_{uv}$, $h^*_{AB_1}=h_u$, and $h^*_{CD}=h_v$, hence
\[
\Delta^*_{CR}=2500-250>0,\qquad \Delta^*_{CC}=50-250<0,\qquad \Delta^*_{MX}=2500-50>0,
\]
which is CRP--RCCP--MXP.

\section{WS and benchmark non-EU predictions}
\label{app:ws-non-eu}

\subsection{Setup: the WS family}
Fix $m>0$. For each $w\in[0,m)$, define the two lotteries
$$
L_1(w) = (w,0.9;\ m,0.05;\ 2m,0.05),
\qquad
L_2(w) = (w,0.9;\ 1.5m,0.1)
$$
as a generalized form of the motivating example used in the main text. Note that both lotteries have the same expected value, $\mathbb{E}[L_1(w)]=\mathbb{E}[L_2(w)]=0.9w+0.15m$, and they share the same worst outcome $w$ with the same probability $0.9$. For a given model evaluation functional $V(\cdot)$, define the WS value difference
$$
\Delta(w) := V(L_1(w)) - V(L_2(w)).
$$
A WS reversal occurs when $\Delta(w)$ changes sign as $w$ varies. A multiple-switch pattern requires $\Delta(w)$ to change sign at least twice.

\subsection{Disappointment aversion (Gul, 1991)}

\paragraph{Model.}
Let $(u,\beta)$ be a Gul (1991) disappointment-aversion (DA) preference, where $u$ is strictly increasing and continuous and $\beta>-1$. For any lottery $p$, let $CE(p)$ denote its DA certainty equivalent in prize units. Let $(a,q,r)$ be the (essentially unique) elation-disappointment decomposition (EDD) of $p$ relative to $CE(p)$: $p=a q + (1-a) r$, all prizes in $q$ are strictly above $CE(p)$, and all prizes in $r$ are strictly below $CE(p)$. The DA value of $p$ in $u$-units is
$$
V_{DA}(p)=\gamma(a)\, \mathbb{E}_q[u(x)] + (1-\gamma(a))\, \mathbb{E}_r[u(x)],
\qquad
\gamma(a)=\frac{a}{1+\beta(1-a)}.
$$
The DA certainty equivalent satisfies $u(CE(p))=V_{DA}(p)$.

\paragraph{Lemma (no reversal while the WS keeps the elation set unchanged).}
Fix $m>0$ and a DA preference $(u,\beta)$. For each $w\in[0,m)$, consider $L_1(w)$ and $L_2(w)$ defined above. Suppose that $CE(L_1(w))<m$. Then the ranking between $L_1(w)$ and $L_2(w)$ is independent of $w$ (equivalently, $\Delta(w)$ is constant in $w$ on this region).

\paragraph{Proof.}
If $CE(L_1(w))<m$, then in $L_1(w)$ the disappointment set is $\{w\}$ and the elation set is $\{m,2m\}$, so the elation probability is $a_1=0.1$ and the EDD is $(0.1,(m,0.5;2m,0.5),\delta_w)$. For $L_2(w)$, we always have $CE(L_2(w))<1.5m$ for $w<m$, so the elation set is $\{1.5m\}$, the disappointment set is $\{w\}$, the elation probability is $a_2=0.1$, and the EDD is $(0.1,\delta_{1.5m},\delta_w)$. Therefore,
$$
V_{DA}(L_1(w))=\gamma(0.1)\frac{u(m)+u(2m)}{2} + (1-\gamma(0.1))u(w),
$$
$$
V_{DA}(L_2(w))=\gamma(0.1)u(1.5m) + (1-\gamma(0.1))u(w),
$$
and hence
$$
\Delta(w)=\gamma(0.1)\left(\frac{u(m)+u(2m)}{2}-u(1.5m)\right),
$$
which is independent of $w$.
\hfill$\square$

\paragraph{Corollary (a sufficient condition in terms of the WS parameters).}
If $(u,\beta)$ is risk-averse in the DA sense (in particular, $u$ is concave and $\beta\ge 0$), then for all $w<\frac{17}{18}m$ we have $CE(L_1(w))<m$, and thus $\Delta(w)$ is constant on $w\in[0,\frac{17}{18}m)$.

\paragraph{Proof.}
When $\beta\ge 0$, we have $\gamma(a)\le a$ for all $a\in(0,1)$, so DA weakly downweights elation relative to EU. With concave $u$, this implies $CE(L_1(w))\le \mathbb{E}[L_1(w)]=0.9w+0.15m$. If $w<\frac{17}{18}m$, then $0.9w+0.15m<m$, so $CE(L_1(w))<m$ and the Lemma applies.
\hfill$\square$

\paragraph{Proposition (single crossing in the WS problem under DA).}
Fix $m>0$ and a DA preference $(u,\beta)$. For $w\in[0,m)$, let $\Delta(w)=V_{DA}(L_1(w))-V_{DA}(L_2(w))$. Then $\Delta(w)$ can change sign at most once as $w$ increases.

\paragraph{Proof.}
As $w$ increases, $L_1(w)$ first-order stochastically increases, so its certainty equivalent $CE(L_1(w))$ is (weakly) increasing in $w$. Therefore, the condition $CE(L_1(w))<m$ can fail at most once. If it never fails, then $\Delta(w)$ is constant by the Lemma and there is no sign change.

If it fails, then there exists $\bar w\in(0,m)$ such that $CE(L_1(w))<m$ for $w<\bar w$ and $CE(L_1(w))\ge m$ for $w\ge \bar w$. For $w\ge \bar w$, the elation set of $L_1(w)$ is $\{2m\}$ and the disappointment set is $\{w,m\}$, so the elation probability is $0.05$ and the EDD is $(0.05,\delta_{2m},(w,\frac{18}{19};m,\frac{1}{19}))$. Thus,
$$
V_{DA}(L_1(w))=\gamma(0.05)u(2m) + (1-\gamma(0.05))\left(\frac{18}{19}u(w)+\frac{1}{19}u(m)\right).
$$
For $L_2(w)$ we always have elation probability $0.1$ and disappointment prize $w$, so
$$
V_{DA}(L_2(w))=\gamma(0.1)u(1.5m) + (1-\gamma(0.1))u(w).
$$
Hence, for $w\ge \bar w$,
$$
\Delta(w)=A\,u(w)+B,
$$
where $A=\left(1-\gamma(0.05)\right)\frac{18}{19}-(1-\gamma(0.1))$ and $B=\gamma(0.05)u(2m)+\left(1-\gamma(0.05)\right)\frac{1}{19}u(m)-\gamma(0.1)u(1.5m)$ are constants in $w$. Since $u$ is strictly increasing, $u(w)$ is strictly increasing in $w$, so $\Delta(w)$ is monotone on $[\bar w,m)$. Combining constancy on $[0,\bar w)$ with monotonicity on $[\bar w,m)$ implies at most one sign change overall.
\hfill$\square$

\subsubsection{Generalized disappointment aversion \autocite{routledge2010}}
We record a closely related implication for the generalized disappointment aversion (GDA) model of
\textcite{routledge2010}. The result mirrors the DA single-crossing logic above: in the WS family, the
GDA disappointment set can change at most once, which implies at most one sign change.

\paragraph{Model.}
Fix a strictly increasing, continuous $u$ and parameters $\theta\ge 0$ and $\delta\in(0,1]$.
For any lottery $p$ over prizes, define its (GDA) certainty equivalent $CE_{GDA}(p)$ as the unique
solution $c$ to
\begin{equation}\label{eq:gda-def}
u(c)
=
\mathbb{E}_p[u(x)]
-\theta\,\mathbb{E}_p\!\left[\bigl(u(\delta c)-u(x)\bigr)\,\mathbf{1}\{x\le \delta c\}\right].
\end{equation}
Define the GDA value functional by $V_{GDA}(p):=u(CE_{GDA}(p))$ and, for the WS lotteries
$L_1(w)$ and $L_2(w)$, define $\Delta_{GDA}(w):=V_{GDA}(L_1(w))-V_{GDA}(L_2(w))$.

\paragraph{Single crossing under homothetic utility.}
Assume $u$ is homothetic in the sense that $u(\delta x)=\delta^{\eta}u(x)$ for some $\eta>0$
(e.g., $u(x)=x^{\eta}$ up to affine normalization). Then $\Delta_{GDA}(w)$ can change sign at most
once as $w$ increases on $[0,m)$.

\paragraph{Proof.}
Under $u(\delta x)=\delta^{\eta}u(x)$ we have $u(\delta c)=\delta^{\eta}u(c)$ for any $c$.
Substituting into \eqref{eq:gda-def} and rearranging yields the following closed-form expression
conditional on the endogenous disappointment set $D(p):=\{x\in\supp(p): x\le \delta\,CE_{GDA}(p)\}$:
\begin{equation}\label{eq:gda-linear}
V_{GDA}(p)
=
\frac{\mathbb{E}_p[u(x)] + \theta\,\mathbb{E}_p[u(x)\mathbf{1}\{x\in D(p)\}]}
{1+\theta\,\delta^{\eta}\,p(D(p))}.
\end{equation}

We now apply \eqref{eq:gda-linear} to the WS family. For $L_2(w)=(w,0.9;1.5m,0.1)$ with $w<m$,
monotonicity implies $CE_{GDA}(L_2(w))\in(w,1.5m)$, hence $\delta\,CE_{GDA}(L_2(w))<1.5m$ (since
$\delta\le 1$). Therefore $1.5m\notin D(L_2(w))$, while $w\in D(L_2(w))$ by definition of $D(\cdot)$,
so $D(L_2(w))=\{w\}$ and $p(D(L_2(w)))=0.9$ for all $w<m$.

For $L_1(w)=(w,0.9;m,0.05;2m,0.05)$, we similarly have $CE_{GDA}(L_1(w))\in(w,2m)$, so
$\delta\,CE_{GDA}(L_1(w))<2m$ and hence $2m\notin D(L_1(w))$ for all $w<m$. Consequently, the only
possible change in $D(L_1(w))$ as $w$ varies is whether $m$ becomes disappointing. Define the cutoff
\[
\bar w:=\inf\{w\in[0,m): m\in D(L_1(w))\},
\]
with $\bar w=m$ if the set is empty. Then:
\begin{itemize}
\item For $w<\bar w$, $D(L_1(w))=\{w\}$ and $p(D(L_1(w)))=0.9$. Since $D(L_2(w))=\{w\}$ as well,
the denominators in \eqref{eq:gda-linear} coincide for $L_1(w)$ and $L_2(w)$ and the $u(w)$ terms
cancel in the difference. Hence $\Delta_{GDA}(w)$ is constant in $w$ on $[0,\bar w)$.
\item For $w\ge \bar w$, $D(L_1(w))=\{w,m\}$ and $p(D(L_1(w)))=0.95$, while $D(L_2(w))=\{w\}$ and
$p(D(L_2(w)))=0.9$. Using \eqref{eq:gda-linear}, $\Delta_{GDA}(w)$ can be written as
$\Delta_{GDA}(w)=A\,u(w)+B$ on $[\bar w,m)$ for constants $A,B$ that do not depend on $w$, with
\[
A=0.9(1+\theta)\left(\frac{1}{1+0.95\,\theta\,\delta^{\eta}}-\frac{1}{1+0.9\,\theta\,\delta^{\eta}}\right)\le 0.
\]
Since $u(w)$ is strictly increasing, $\Delta_{GDA}(w)$ is weakly decreasing on $[\bar w,m)$.
\end{itemize}
Therefore $\Delta_{GDA}(w)$ is constant on $[0,\bar w)$ and weakly decreasing on $[\bar w,m)$, so it
can cross zero at most once on $[0,m)$. \hfill$\square$

\paragraph{A sufficient WS range with no change in the GDA disappointment set.}
Assume \eqref{eq:gda-def} with $\theta\ge 0$, $\delta\in(0,1]$, and concave $u$. Then for all
\[
w < \frac{m}{0.9}\Big(\frac{1}{\delta}-0.15\Big),
\]
we have $\delta\,CE_{GDA}(L_1(w))<m$, hence $m\notin D(L_1(w))$ and $\Delta_{GDA}(w)$ is constant in
$w$ on this region (no WS reversal can occur there). In particular, when $\delta=1$ this reduces to
$w<\frac{17}{18}m$.

\paragraph{Proof.}
From \eqref{eq:gda-def} and $\theta\ge 0$,
\[
u(CE_{GDA}(p))\le \mathbb{E}_p[u(x)]\quad\Rightarrow\quad CE_{GDA}(p)\le u^{-1}(\mathbb{E}_p[u(x)]).
\]
If $u$ is concave, Jensen implies $u^{-1}(\mathbb{E}_p[u(x)])\le \mathbb{E}_p[x]$, hence
$CE_{GDA}(p)\le \mathbb{E}_p[x]$. For $p=L_1(w)$, $\mathbb{E}[L_1(w)]=0.9w+0.15m$, so the stated
bound on $w$ implies $0.9w+0.15m<m/\delta$, that is, $CE_{GDA}(L_1(w))<m/\delta$ and therefore
$\delta\,CE_{GDA}(L_1(w))<m$. This ensures $m\notin D(L_1(w))$, so we remain in the ``low'' case
$D(L_1(w))=\{w\}$ and $\Delta_{GDA}(w)$ is constant in $w$ as in the first bullet of the proof above.
\hfill$\square$

\subsection{Cautious expected utility (Cerreia-Vioglio, Dillenberger, and Ortoleva, 2015)}

\paragraph{Model.}
Assume a cautious expected utility (CEU) representation
$$
V_{CEU}(p)=\inf_{v\in\mathcal{V}} v^{-1}\left(\mathbb{E}_p[v(x)]\right),
$$
where $\mathcal{V}$ is a nonempty set of strictly increasing functions on the prize interval. (The object $v^{-1}(\mathbb{E}_p[v])$ is the certainty equivalent of $p$ under $v$.)

\paragraph{Proposition (no WS reversal under CEU with uniformly concave or uniformly convex $\mathcal{V}$).}
Fix $m>0$ and $w\in[0,m)$. If every $v\in\mathcal{V}$ is concave, then $V_{CEU}(L_1(w))\le V_{CEU}(L_2(w))$ for all $w<m$. If every $v\in\mathcal{V}$ is convex, then $V_{CEU}(L_1(w))\ge V_{CEU}(L_2(w))$ for all $w<m$. In either case, the ranking is independent of $w$ and CEU rules out WS reversals (and hence multiple switches) in this WS family.

\paragraph{Proof.}
Fix any $v\in\mathcal{V}$. Since $1.5m=\frac12 m+\frac12(2m)$, concavity of $v$ implies
$$
\frac12 v(m)+\frac12 v(2m)\le v(1.5m),
$$
or equivalently $0.05v(m)+0.05v(2m)\le 0.1v(1.5m)$. Therefore,
$$
\mathbb{E}[v(L_1(w))]-\mathbb{E}[v(L_2(w))]=0.05v(m)+0.05v(2m)-0.1v(1.5m)\le 0,
$$
where the common $0.9v(w)$ term cancels. Since $v^{-1}$ is increasing, this yields
$$
v^{-1}(\mathbb{E}[v(L_1(w))])\le v^{-1}(\mathbb{E}[v(L_2(w))]).
$$
Taking $\inf_{v\in\mathcal{V}}$ preserves the inequality, so $V_{CEU}(L_1(w))\le V_{CEU}(L_2(w))$ for all $w<m$. The convex case is identical with the inequality reversed by Jensen.
\hfill$\square$

\subsection{Upside potential (McGranaghan et al., 2025)}

\paragraph{Model definition}
Let $W$ be a set of outcomes considered to be ``winning outcomes.'' The upside-potential (UP) model evaluates a lottery $X=(x_1,p_1;\ldots;x_N,p_N)$ as
$$
U_{UP}(X)=\sum_{i=1}^N p_i u(x_i) + \left(\sum_{j=1}^N p_j \mathbf{1}\{x_j\in W\}\right)\left(\sum_{k=1}^N p_k \mathbf{1}\{x_k\in W\}\kappa(x_k)\right),
$$
where $u$ and $\kappa$ are strictly increasing and $\kappa(x)\ge 0$ for $x\in W$. In their application with zero versus positive payoffs, a maintained assumption is $W=\{x>0\}$ (all positive outcomes are winning, zero is not).

\paragraph{Proposition (UP implies at most one WS switch at the win-threshold).}
Fix $m>0$ and define $L_1(w),L_2(w)$ as above. Assume the UP model with a threshold-type winning set $W=\{x>\bar d\}$ for some $\bar d<m$ (so that $m,1.5m,2m\in W$). Then the UP value difference
$$
\Delta_{UP}(w):=U_{UP}(L_1(w))-U_{UP}(L_2(w))
$$
is constant in $w$ for $w\le \bar d$ and constant in $w$ for $w>\bar d$. Consequently, $\Delta_{UP}(w)$ can change sign at most once as $w$ increases, and multiple switches are ruled out in this WS family.

\paragraph{Proof.}
If $w\le \bar d$, then $w\notin W$ while $m,1.5m,2m\in W$. The EU components share the common $0.9u(w)$ term, so the $w$-dependence cancels in $\Delta_{UP}(w)$. The UP terms also exclude $\kappa(w)$ because $w\notin W$, and both lotteries have total winning probability $0.1$, so $\Delta_{UP}(w)$ is constant for all $w\le \bar d$.

If $w>\bar d$, then all outcomes in both lotteries lie in $W$, so $\sum p_j \mathbf{1}\{x_j\in W\}=1$ for both lotteries and the UP term becomes $\sum p_k \kappa(x_k)$. Since both lotteries share the outcome $w$ with probability $0.9$, the $0.9\kappa(w)$ terms cancel as well, and $\Delta_{UP}(w)$ is constant for all $w>\bar d$. The only possible change in $\Delta_{UP}(w)$ occurs at the unique boundary $w=\bar d$, so at most one sign change can occur.
\hfill$\square$

\paragraph{Special case ($W=\{x>0\}$).}
If $W=\{x>0\}$ and $w\in[0,m)$, the only boundary is at $w=0$. Therefore, the UP model predicts that the WS ranking is invariant in $w$ on $(0,m)$, with a possible discrete change when the worst outcome moves from $w=0$ to $w>0$.

\section{Feldman--Rehbeck overlay methods and full results}\label{app:feldman-rehbeck-overlay}

\subsection{Provenance and scope}\label{app:feldman-rehbeck-overlay:provenance}
We apply the ECU overlay to the convex-choice data of Feldman and Rehbeck (2022), using their replication package as the source of truth for the task design and observed mixing weights $\alpha_{\mathrm{obs}}$.

\subsection{Convex-choice environment and notation}\label{app:feldman-rehbeck-overlay:environment}
The dataset contains $N_{\mathrm{subj}}=144$ subjects and $T=79$ convex tasks per subject, for $N_{\mathrm{subj}}\times T=11{,}376$ observed mixing choices. In each convex task, the subject chooses a mixing weight $\alpha\in[0,100]$ between two fixed endpoint lotteries over outcomes $x\in\{2,10,30\}$, which we denote by the ``Extreme'' endpoint ($\alpha=0$) and the ``Numeraire'' endpoint ($\alpha=100$).

Let $w=\alpha/100$. Denote the endpoint probabilities of outcome $x\in\{2,10,30\}$ by $p_x(0)$ and $p_x(100)$, and write the mixed lottery induced by $\alpha$ as
$$
p_x(\alpha)=(1-w)p_x(0)+wp_x(100),\qquad x\in\{2,10,30\}.
$$
For convenience, let $p_C(\alpha)=p_{2}(\alpha)$, $p_B(\alpha)=p_{10}(\alpha)$, and $p_A(\alpha)=p_{30}(\alpha)$.

\subsection{Models}\label{app:feldman-rehbeck-overlay:models}

\paragraph{EU benchmark.}\label{app:feldman-rehbeck-overlay:eu}
The EU benchmark uses the normalized utility $u(2)=0$, $u(30)=1$, and a single free parameter $u(10)=m$. The predicted choice is
$$
\alpha_{\mathrm{pred}}^{\mathrm{EU}}\in\arg\max_{\alpha\in[0,100]} \Big(p_A(\alpha)\cdot 1+p_B(\alpha)\cdot m+p_C(\alpha)\cdot 0\Big).
$$
Because $p_x(\alpha)$ is linear in $\alpha$, the EU objective is linear in $\alpha$, hence $\alpha_{\mathrm{pred}}^{\mathrm{EU}}$ is (generically) an endpoint in $\{0,100\}$.

\paragraph{ECU overlay (two contexts).}\label{app:feldman-rehbeck-overlay:ecu2}
The ECU overlay keeps the same normalization $u(2)=0$ and $u(30)=1$, but allows the utility of the middle outcome to depend on the lottery's context, summarized by a disappointment probability $\pi(\alpha;d)$. The ``disappointing set'' is $D=\{x\le d\}$, and we consider $d\in\{2,10\}$, where $d=2$ is the main specification and $d=10$ is a robustness alternative:
$$
\pi(\alpha;2)=\Pr[x\le 2]=p_C(\alpha),\qquad \pi(\alpha;10)=\Pr[x\le 10]=p_C(\alpha)+p_B(\alpha)=1-p_A(\alpha).
$$
The ECU context rule is a two-context step function:
$$
m(\pi)=
\begin{cases}
m_{\mathrm{low}} & \text{if }\pi\le \tau,\\
m_{\mathrm{high}} & \text{if }\pi>\tau,
\end{cases}
$$
so that $u(10)=m(\pi(\alpha;d))$. Given parameters $(m_{\mathrm{low}},m_{\mathrm{high}},\tau)$, the ECU value of choosing $\alpha$ is
$$
V(\alpha)=p_A(\alpha)\cdot 1+p_B(\alpha)\cdot m(\pi(\alpha;d))+p_C(\alpha)\cdot 0,
$$
and the predicted choice is
$$
\alpha_{\mathrm{pred}}^{\mathrm{ECU}}\in\arg\max_{\alpha\in[0,100]} V(\alpha).
$$
Since $\pi(\alpha;d)$ is linear in $\alpha$, $V(\alpha)$ is piecewise linear with at most one kink at $\alpha_{\mathrm{kink}}$ solving $\pi(\alpha_{\mathrm{kink}};d)=\tau$. Therefore, we compute $\alpha_{\mathrm{pred}}^{\mathrm{ECU}}$ by evaluating $V(\alpha)$ on the candidate set $\{0,100,\alpha_{\mathrm{kink}}\}$ whenever $\alpha_{\mathrm{kink}}\in(0,100)$, and use the tie-breaking rule ``choose the smallest $\alpha$''.

\paragraph{ECU overlay (three contexts, robustness).}\label{app:feldman-rehbeck-overlay:ecu3}
As an additional robustness check, we allow a three-context step function
$$
m(\pi)=
\begin{cases}
m_1 & \text{if }\pi\le \tau_1,\\
m_2 & \text{if }\tau_1<\pi\le \tau_2,\\
m_3 & \text{if }\pi>\tau_2,
\end{cases}
$$
with constraints $m_1\le m_2\le m_3$ and $\tau_1\le \tau_2$. With three contexts, $V(\alpha)$ is piecewise linear with at most two kinks; we evaluate $V(\alpha)$ on the candidate set $\{0,100,\alpha_{\mathrm{k1}},\alpha_{\mathrm{k2}}\}$ for kinks in $(0,100)$, again breaking ties by choosing the smallest $\alpha$.

\subsection{Distance metric and estimation}\label{app:feldman-rehbeck-overlay:estimation}
For each subject $i$ and task $t$, let $\alpha_{\mathrm{obs},it}$ be the observed mixing weight and let $\alpha_{\mathrm{pred},it}(\theta)$ be the predicted weight under parameter vector $\theta$. The row-level distance is $|\alpha_{\mathrm{obs},it}-\alpha_{\mathrm{pred},it}(\theta)|$, and the primary fit criterion is the subject-weighted mean absolute distance
$$
L(\theta)=\frac{1}{N_{\mathrm{subj}}}\sum_{i=1}^{N_{\mathrm{subj}}}\left(\frac{1}{T}\sum_{t=1}^{T}\left|\alpha_{\mathrm{obs},it}-\alpha_{\mathrm{pred},it}(\theta)\right|\right).
$$
We estimate parameters by grid search. For EU we search $m\in[0.05,0.95]$ on a fine grid with step size $0.01$. For the two-context ECU overlay we search $m_{\mathrm{low}},m_{\mathrm{high}}\in[0.05,0.95]$ and $\tau\in[0.05,0.95]$ on a coarser grid with step size $0.05$. This makes the one-parameter EU benchmark conservative relative to ECU in terms of grid granularity. For the three-context robustness extension we search $m_1,m_2,m_3\in[0.05,0.95]$ and $\tau_1,\tau_2\in[0.05,0.95]$ on the same $0.05$ grid, imposing $m_1\le m_2\le m_3$ and $\tau_1\le\tau_2$.

In addition to $L(\theta)$, we report the share of subjects better fit by ECU than EU, defined as the fraction of subjects for whom the subject-level mean distance is smaller under ECU than under EU.

\subsection{Results}\label{app:feldman-rehbeck-overlay:results}

\paragraph{Two-context ECU (main and robustness).}\label{app:feldman-rehbeck-overlay:results-2reg}
Table \ref{tab:ecu-overlay-2reg} reports the best-fit parameters and the minimized subject-weighted mean absolute distance. Relative to the one-parameter EU benchmark, the two-context ECU overlay improves fit modestly in both disappointment definitions ($d=2$ and $d=10$) and improves the fit for a majority of subjects.

\begin{table}[!htbp]\centering
\caption{Best-fit EU and two-context ECU overlay (subject-weighted mean absolute distance)}\label{tab:ecu-overlay-2reg}
\begin{tabular}{l c l c c}
\hline
Model & $d$ & Parameters & $L(\hat\theta)$ & Share of subjects improved \\
\hline
EU & N/A & $m=0.66$ & 29.00 & N/A \\
ECU (two contexts) & 2 & $m_{\mathrm{low}}=0.65,\ m_{\mathrm{high}}=0.85,\ \tau=0.05$ & 28.68 & 58.3\% \\
ECU (two contexts) & 10 & $m_{\mathrm{low}}=0.65,\ m_{\mathrm{high}}=0.60,\ \tau=0.85$ & 28.59 & 61.1\% \\
\hline
\end{tabular}
\end{table}

\paragraph{Three-context ECU robustness.}\label{app:feldman-rehbeck-overlay:results-3reg}
Table \ref{tab:ecu-overlay-3reg} reports the three-context robustness results. For the main disappointment definition $d=2$, allowing a second threshold yields a larger improvement in fit relative to the two-context ECU overlay. For $d=10$, the three-context extension does not improve on the two-context overlay.

\begin{table}[!htbp]\centering
\caption{Three-context ECU robustness (subject-weighted mean absolute distance)}\label{tab:ecu-overlay-3reg}
\begin{tabular}{l c l c c}
\hline
Model & $d$ & Parameters & $L(\hat\theta)$ & Share of subjects improved \\
\hline
EU & N/A & $m=0.66$ & 29.00 & N/A \\
ECU (two contexts) & 2 & $m_{\mathrm{low}}=0.65,\ m_{\mathrm{high}}=0.85,\ \tau=0.05$ & 28.68 & 58.3\% \\
ECU (three contexts) & 2 & $(m_1,m_2,m_3)=(0.65,0.80,0.85),\ (\tau_1,\tau_2)=(0.05,0.10)$ & 26.58 & 63.9\% \\
ECU (two contexts) & 10 & $m_{\mathrm{low}}=0.65,\ m_{\mathrm{high}}=0.60,\ \tau=0.85$ & 28.59 & 61.1\% \\
ECU (three contexts) & 10 & $(m_1,m_2,m_3)=(0.05,0.65,0.75),\ (\tau_1,\tau_2)=(0.05,0.90)$ & 28.78 & 50.7\% \\
\hline
\end{tabular}
\end{table}

\paragraph{Supplementary outputs (row-level and by-budget summaries).}\label{app:feldman-rehbeck-overlay:outputs}
The overlay code also produces row-level predictions for each task and subject (including $\alpha_{\mathrm{pred}}$ and row-level distances), summary statistics computed both pooled and subject-weighted, and fit summaries by the canonical Feldman--Rehbeck budget subsets (N1--N7), along with a binned ``improvement versus disappointment probability'' dataset intended for plotting in the main text.

\section{Stage 3 construction (Crossing vs Non-Crossing)}
\label{app:stage3-construction}

This appendix documents the construction of the Stage 3 choice pairs used to link the Stage 1 and Stage 2 switching locations to Allais-type reversals. The main text describes Stage 3 as a robustness exercise: using participant-specific switching locations from Stages 1 and 2 (interpreted under the one-switch protocol as $(\hat d_i,\hat\tau_i)$ in the binary ECU benchmark), we construct standard common-consequence (CC) and common-ratio (CR) comparisons that are designed either to cross the inferred threshold partition (the \emph{Crossing} set) or to remain within a single region (the \emph{Non-Crossing} set).

\subsection{Notation and implementation}
Fix a prize level $b$ (the high outcome used in the Stage 3 blocks). For each participant $i$, let $\hat d_i$ denote the Stage 1 switching location (the recovered disappointment threshold under the one-switch protocol) and let $\hat\tau_i$ denote the Stage 2 switching location (the recovered tolerance threshold). For compactness, throughout this appendix write $d=\hat d_i$ and $\tau=\hat\tau_i$. We also fix $\epsilon=0.01$.

We write lotteries in the form $(x,p;y,q;z,r)$, where outcomes $x,y,z$ occur with probabilities $p,q,r$ and probabilities sum to one (terms omitted when a probability is zero). Stage 3 tasks are presented one at a time (rather than in a list) to reduce the scope for induced preferences for randomization (see \textcite{Agranov2017} and \textcite{Feldman2022}).

\subsection{Crossing set}
The Crossing set consists of CC and CR blocks in which at least one of the relevant options is constructed to cross the inferred partition when moving from the first task to the second task within the block.

The Crossing--CC pair is:
$$
\left(d+\frac{b-d}{2},1\right)\ \text{vs.}\ \left(d+\frac{3b-3d}{4},\frac{1-\tau-\epsilon}{2};\ d+\frac{b-d}{2},\tau+\epsilon;\ 0,\frac{1-\tau-\epsilon}{2}\right),
$$
$$
\left(d+\frac{b-d}{2},1-\tau-\epsilon;\ 0,\tau+\epsilon\right)\ \text{vs.}\ \left(d+\frac{3b-3d}{4},\frac{1-\tau-\epsilon}{2};\ 0,1-\frac{1-\tau-\epsilon}{2}\right).
$$
Intuition: the first task features a CC-style comparison with a shared intermediate outcome $d+\frac{b-d}{2}$. The second task is obtained by transferring probability mass $\tau+\epsilon$ from that shared outcome to the worst outcome $0$. This transfer makes it possible (under the binary ECU benchmark) for the relevant option to be evaluated under a different context after the transfer, which is the mechanism that can generate an Allais-type reversal within the block.

The Crossing--CR pair is:
$$
(b,0.8;\ 0,0.2)\ \text{vs.}\ \left(d+\frac{b-d}{2},1\right),
$$
$$
(b,0.8\times(1-\tau-\epsilon);\ 0,1-0.8\times(1-\tau-\epsilon))\ \text{vs.}\ \left(d+\frac{b-d}{2},1-\tau-\epsilon;\ 0,\tau+\epsilon\right).
$$
Intuition: moving from the first task to the second task applies a common-ratio style scaling on the left lottery while simultaneously constructing the right lottery so that the probability on $0$ rises to $\tau+\epsilon$. This is designed to create a context change within the block in a way that is parallel to the classical CR comparison.

\subsection{Non-Crossing set}
The Non-Crossing set consists of CC and CR blocks constructed so that the relevant options are designed to remain within a single context across the two tasks in the block (so that the binary ECU benchmark does not generate an Allais-type reversal via a context change within the block).

The Non-Crossing--CC pair is:
$$
\left(\frac{d}{10}+\frac{b+d}{2},1\right)\ \text{vs.}\ \left(\frac{d}{5}+\frac{b+d}{2},0.5;\ \frac{d}{10}+\frac{b+d}{2},0.1;\ 0,0.4\right),
$$
$$
\left(\frac{d}{10}+\frac{b+d}{2},0.9;\ 0,0.1\right)\ \text{vs.}\ \left(\frac{d}{5}+\frac{b+d}{2},0.5;\ 0,0.5\right).
$$

The Non-Crossing--CR pair is:
$$
\left(\frac{d+\epsilon}{4}+\frac{b+d}{2},1\right)\ \text{vs.}\ \left(\frac{d+\epsilon}{3}+\frac{b+d}{2},0.5;\ 0,0.5\right),
$$
$$
\left(\frac{d+\epsilon}{4}+\frac{b+d}{2},0.9;\ 0,0.1\right)\ \text{vs.}\ \left(\frac{d+\epsilon}{3}+\frac{b+d}{2},0.45;\ 0,0.55\right).
$$

\noindent We note that for participants with $\hat\tau = 0.4$ (CC block) or $\hat\tau = 0.5$ (CR block), the right-side option in the Non-Crossing block can cross the inferred context boundary. This contamination attenuates rather than inflates the Crossing vs. Non-Crossing contrast, making our reported effect conservative for these participants.

\section{Additional experimental results}
\label{app:additional-experimental-results}

This appendix collects additional descriptive results that complement the main regularities reported in the paper. We focus on (i) early laboratory sessions that allowed unrestricted switching (to document multiple switching and motivate the one-switch protocol), (ii) a summary of laboratory sessions implementing the one-switch protocol (reported as a separate environment/condition), (iii) exploratory heterogeneity by self-reported gender in the online session, and (iv) supplementary test statistics reported for completeness.

\subsection{Pilot laboratory sessions with unrestricted switching (multiple switching)}
\label{app:unrestricted-switching}

Before implementing the one-switch protocol used in the main sessions, we conducted laboratory pilot sessions in which participants could switch freely between Option A and Option B across rows. In these pilots, multiple switching is common. This is informative for two reasons. First, it provides a direct descriptive link to multiple switching behavior (MSB) in choice lists. Second, it complicates the simple recovery of $(d,\tau)$ under the binary ECU benchmark, which motivates restricting switching to at most one in the main sessions.

\paragraph{Pilot session (unrestricted switching).} In a first pilot session ($n=14$), 10 of 14 participants (71.4\%) switched at least once in Stage 1, and conditional on switching at least once, participants switched 4.1 times on average. In Stage 2, 12 of 14 participants (85.7\%) switched at least once, and conditional on switching at least once, participants switched 3.25 times on average. Switching was related across stages: 9 of 10 Stage 1 switchers (90.0\%) also switched in Stage 2; among the 4 Stage 1 non-switchers, 3 (75.0\%) switched in Stage 2.

\paragraph{High-stakes pilot session (unrestricted switching).} To assess whether higher stakes reduce excessive switching, we ran a second pilot session ($n=14$) with fourfold prizes (earnings ranged from \$7 to \$103 for the 30-minute session). In Stage 1, 11 of 14 participants (78.6\%) switched at least once; conditional on switching at least once, they switched 2.55 times on average. In Stage 2, 13 of 14 participants (92.9\%) switched at least once; conditional on switching at least once, they switched 3.46 times on average. Among Stage 1 switchers, 10 of 11 (90.9\%) also switched in Stage 2, and all 3 Stage 1 non-switchers switched in Stage 2. Relative to the first pilot, higher stakes increased the fraction of participants who switched only once and reduced the average number of Stage 1 switches, though the small sample size limits strong conclusions.

\subsection{Laboratory sessions with the one-switch protocol}
\label{app:lab-one-switch}

In later laboratory sessions we implemented the one-switch protocol used in the main analysis. The patterns are qualitatively consistent with the main regularities.

\paragraph{Stages 1 and 2.} In these laboratory sessions ($n=67$), 42 of 67 participants (62.7\%) switched across the ten-row menu in Stage 1. In Stage 2, 49 of 67 participants (73.1\%) switched. Switching is strongly related across stages: 37 of 42 Stage 1 switchers (88.1\%) also switched in Stage 2, whereas 12 of 25 Stage 1 non-switchers (48.0\%) switched in Stage 2.

\paragraph{Stage 3.} In the set of Stage 3 tasks designed to elicit Allais-type reversals under the ECU construction, 13 of 67 participants (19.4\%) reversed in the CC block and 21 of 67 (31.3\%) reversed in the CR block. In the set designed to not elicit such reversals, 3 of 67 (4.5\%) reversed in the CC block and 0 of 67 (0\%) reversed in the CR block.

\subsection{Exploratory heterogeneity by self-reported gender (online session)}
\label{app:gender-heterogeneity}

In the online session, we examine exploratory heterogeneity by self-reported gender. Table~\ref{tab:gender-stage1-stage2} reports cross-tabulations for Stage 1 and Stage 2 switching. Female participants were less likely to switch in Stage 1. Stage 2 shows a similar direction. However, when conditioning on Stage 1 switchers (Table~\ref{tab:gender-stage2-conditional}), the gender difference in Stage 2 switching is no longer evident. Interpreted through the ECU lens, this suggests that gender may be correlated with parameters affecting Stage 1 switching, and conditioning on the recovered Stage 1 behavior accounts for part of the raw gender difference in Stage 2.

\begin{table}[!htbp]\centering
\caption{Gender cross-tabulations for Stage 1 and Stage 2 switching (online session)}
\label{tab:gender-stage1-stage2}
\begin{tabular}{lccc|lccc}
\hline\hline
& \multicolumn{3}{c|}{Stage 1} & & \multicolumn{3}{c}{Stage 2} \\
& Female & Male & Total & & Female & Male & Total \\
\hline
No switch & 34 & 38 & 72 & No switch & 27 & 31 & 58 \\
Switch & 24 & 53 & 77 & Switch & 31 & 60 & 91 \\
\hline
Total & 58 & 91 & 149 & Total & 58 & 91 & 149 \\
\hline
\multicolumn{4}{l|}{Fisher exact (two-sided): 0.064;} &
\multicolumn{4}{l}{Fisher exact (two-sided): 0.168;} \\
\multicolumn{4}{l|}{(one-sided): 0.033} &
\multicolumn{4}{l}{(one-sided): 0.088} \\

\hline\hline
\end{tabular}
\end{table}

\begin{table}[!htbp]\centering
\caption{Gender cross-tabulation for Stage 2 switching conditional on switching in Stage 1 (online session)}
\label{tab:gender-stage2-conditional}
\begin{tabular}{lccc}
\hline\hline
& Female & Male & Total \\
\hline
No switch & 6 & 8 & 14 \\
Switch & 18 & 45 & 63 \\
\hline
Total & 24 & 53 & 77 \\
\hline
\multicolumn{4}{l}{Fisher exact (two-sided): 0.345;\ \ (one-sided): 0.231} \\
\hline\hline
\end{tabular}
\end{table}

\subsection{Supplementary exact binomial tests}
\label{app:binomial-tests}

For completeness, Tables~\ref{tab:binom-main-stage1}--\ref{tab:binom-pilot-stage2} report exact binomial test outputs that were used in earlier drafts as compact summaries of whether switch rates exceed 0.5 in selected subsamples.

\begin{table}[!htbp]\centering
\caption{Stage 1 exact binomial test (online main session)}
\label{tab:binom-main-stage1}
\begin{tabular}{lc}
\hline\hline
Number of successes & 78 \\
Number of trials & 150 \\
Probability of success & 0.52 \\
95\% confidence interval & $[0.4497293,\ 1.0000000]$ \\
Alternative & $\Pr(\text{success})>0.5$ \\
$p$-value & 0.3416 \\
\hline\hline
\end{tabular}
\end{table}

\begin{table}[!htbp]\centering
\caption{Stage 2 exact binomial test among Stage 1 switchers (online main session)}
\label{tab:binom-main-stage2}
\begin{tabular}{lc}
\hline\hline
Number of successes & 64 \\
Number of trials & 78 \\
Probability of success & 0.8205 \\
95\% confidence interval & $[0.7337,\ 1.0000]$ \\
Alternative & $\Pr(\text{success})>0.5$ \\
$p$-value & $4.291\times 10^{-9}$ \\
\hline\hline
\end{tabular}
\end{table}

\begin{table}[!htbp]\centering
\caption{Stage 1 exact binomial test (laboratory one-switch sessions)}
\label{tab:binom-pilot-stage1}
\begin{tabular}{lc}
\hline\hline
Number of successes & 42 \\
Number of trials & 67 \\
Probability of success & 0.6269 \\
95\% confidence interval & $[0.5194,\ 1.0000]$ \\
Alternative & $\Pr(\text{success})>0.5$ \\
$p$-value & 0.0249 \\
\hline\hline
\end{tabular}
\end{table}

\begin{table}[!htbp]\centering
\caption{Stage 2 exact binomial test among Stage 1 switchers (laboratory one-switch sessions)}
\label{tab:binom-pilot-stage2}
\begin{tabular}{lc}
\hline\hline
Number of successes & 37 \\
Number of trials & 42 \\
Probability of success & 0.8810 \\
95\% confidence interval & $[0.7658,\ 1.0000]$ \\
Alternative & $\Pr(\text{success})>0.5$ \\
$p$-value & $2.217\times 10^{-7}$ \\
\hline\hline
\end{tabular}
\end{table}

\subsection{Logit regression for Stage 3 reversals (supplementary)}
\label{app:stage3-logit}

As an additional summary, we estimate a logit specification at the task-pair level with standard errors clustered at the participant level. Let $y_{t,i}$ indicate whether participant $i$ exhibits a reversal in task pair $t$. Let $\theta_t$ indicate whether the task pair belongs to the set designed to elicit Allais-type reversals under the ECU construction (the Crossing set in the terminology of the revised draft). We include participant characteristics, including the recovered thresholds $d_i$ and $\tau_i$ and an indicator for male.

$$
y_{t,i}=\begin{cases}
1 & \text{if }\beta_0+\beta_1\theta_t+\beta_2 d_i+\beta_3\tau_i+\beta_4\text{Male}_i+\epsilon_{t,i}>0,\\
0 & \text{otherwise.}
\end{cases}
$$

\begin{table}[!htbp]\centering
\caption{Logit regression results (supplementary)}
\label{tab:stage3-logit}
\begin{tabular}{lcccc}
\hline\hline
& Estimate & Std.\ Error & $z$ value & Pr$(>|z|)$ \\
\hline
(Intercept) & $-1.681$ & $0.473$ & $-3.555$ & $0.000$ \\
$\theta_t$ (Crossing-set indicator) & $1.405$ & $0.363$ & $3.876$ & $0.000$ \\
$d_i$ & $-0.002$ & $0.002$ & $-0.856$ & $0.392$ \\
$\tau_i$ & $0.604$ & $0.827$ & $0.731$ & $0.465$ \\
Male$_i$ & $-0.692$ & $0.418$ & $-1.654$ & $0.098$ \\
\hline\hline
\end{tabular}
\end{table}

\section{Screenshots of the experimental software}
\label{app:screenshots}
    \begin{figure}[H]
        \centering
        \includegraphics[width=0.9\textwidth]{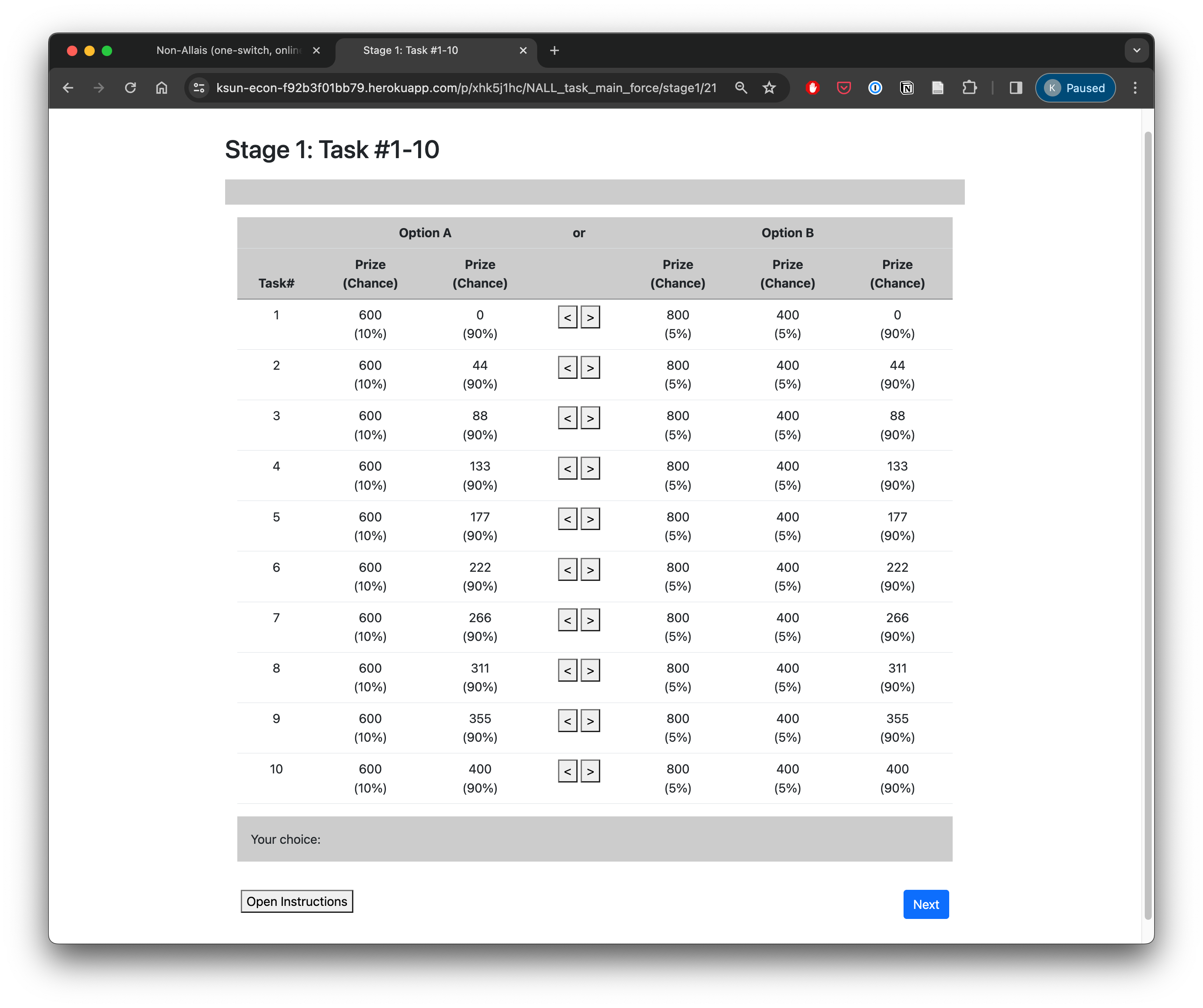}
        \caption{A screenshot of the experiment software displaying Stage 1.}
        \label{fig:stage1_screenshot}
    \end{figure}

    \begin{figure}[H]
        \centering
        \includegraphics[width=0.9\textwidth]{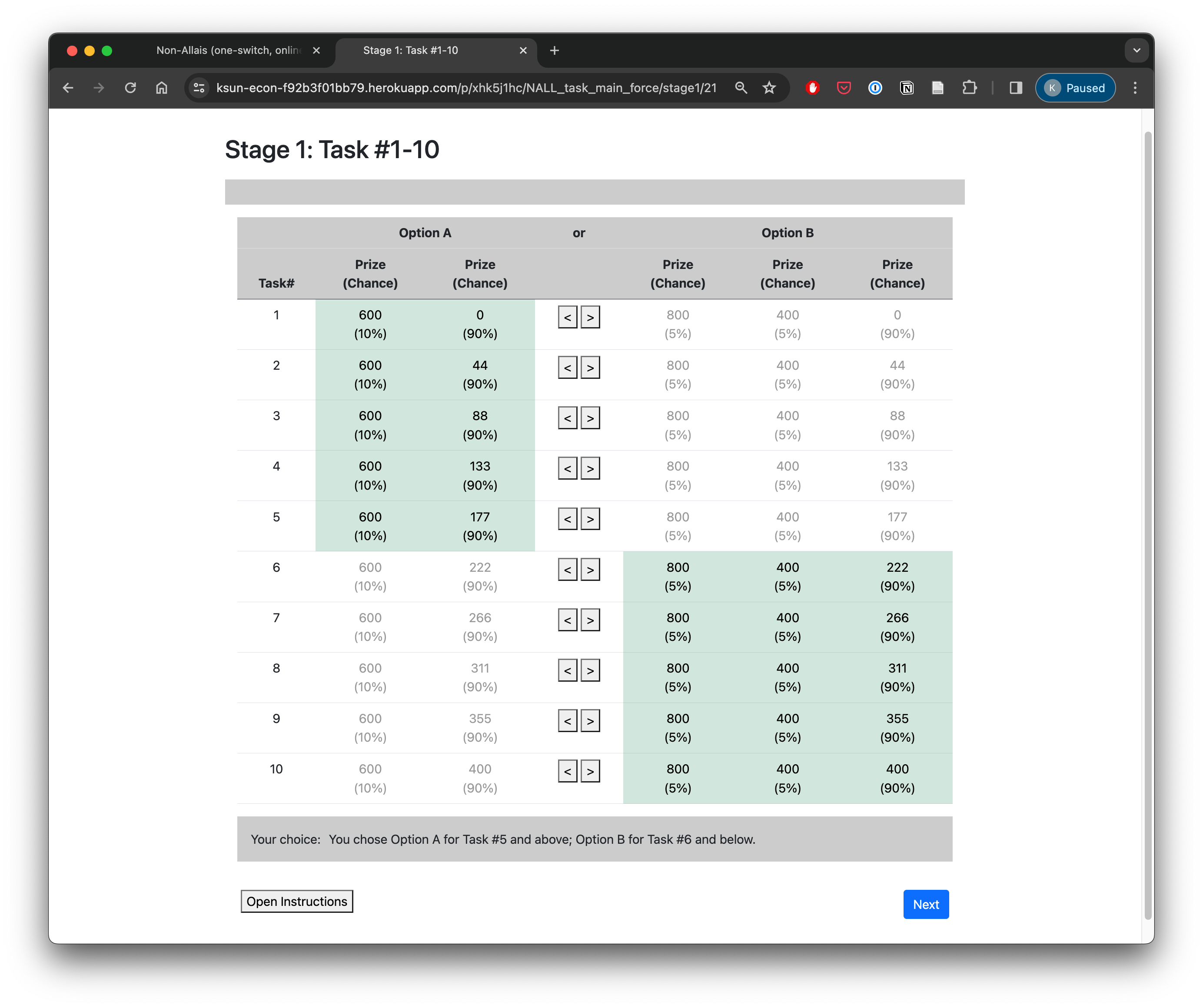}
        \caption{A screenshot of the experiment software displaying Stage 1 where the participant chose Option A for Task \#5 and above.}
        \label{fig:stage1_screenshot_after_choice}
    \end{figure}

    \begin{figure}[H]
        \centering
        \includegraphics[width=0.9\textwidth]{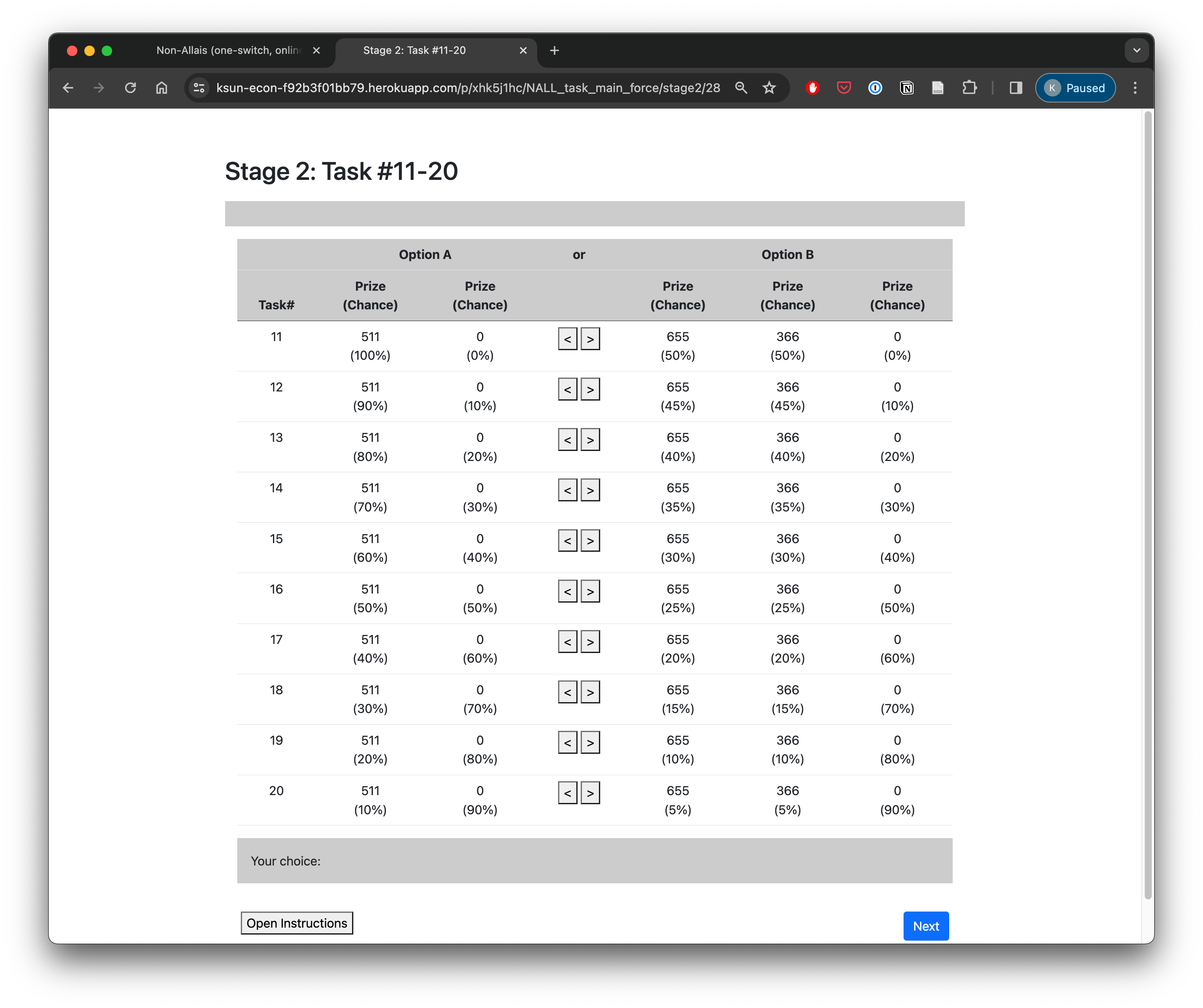}
        \caption{A screenshot of the experiment software displaying Stage 2.}
        \label{fig:stage2_screenshot}
    \end{figure}

    \section{Experimental Instructions}
\label{app:instructions}
Welcome to the experiment! Thank you very much for participating today.
This research is conducted by researchers of the University of Manitoba, Canada and Fairfield University, USA.

The experiment you will be participating in is an experiment in individual decision making.
You will receive your compensation via Prolific.
You will receive the show-up fee of \$6.00 for completing the experiment, with the additional bonus amount that depends on your decisions and on chance.
The details of the compensation will be described later.
All instructions and descriptions that you will be given in the experiment are accurate and true.
At no point will we attempt to deceive you in any way.
When you are ready, please click the button below to read the instructions.

There are three parts in this experiment.

\begin{figure}[ht]
    \centering
    \includegraphics[width=0.9\textwidth]{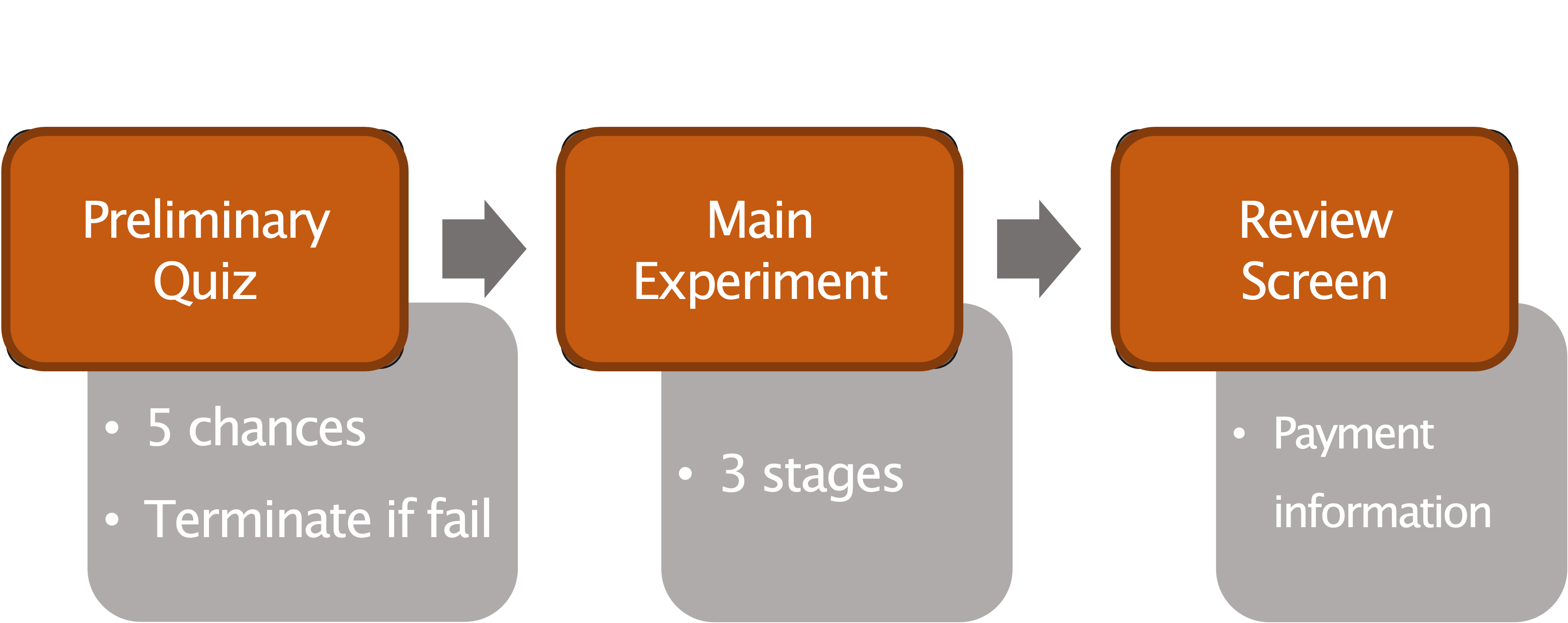}
\end{figure}

First, there is a short quiz at the end of the instructions to ensure your understanding of the procedures.
You will be able to repeat the quiz if you make mistakes.
You will have \textbf{five} chances to attempt the quiz.
If you fail to get all questions correct after five attempts, you may \textbf{not} participate in the main experiment.
If you are unable to pass the quiz, please understand that we can only compensate those who participate in the main part of the experiment and we will not be able to provide payment.
In such an event, we kindly request that you return your submission on Prolific.

In the main part of the experiment, we will ask you to perform tasks of choosing between two options.
We will shortly explain what they are.

At the end of the experiment, we will display a review screen to verify your payment information.

We will now give you more details about the main experiment. The basic unit of the experiment is a stage of tasks. Each stage consists of one or more tasks.

In each stage, we ask you to perform a series of tasks of choosing between two available options.
Each option in a task represents a lottery by which you can earn certain amounts of points depending on chances.
Below is an example of such options.

\begin{figure}[ht]
    \centering
    \includegraphics[width=0.9\textwidth]{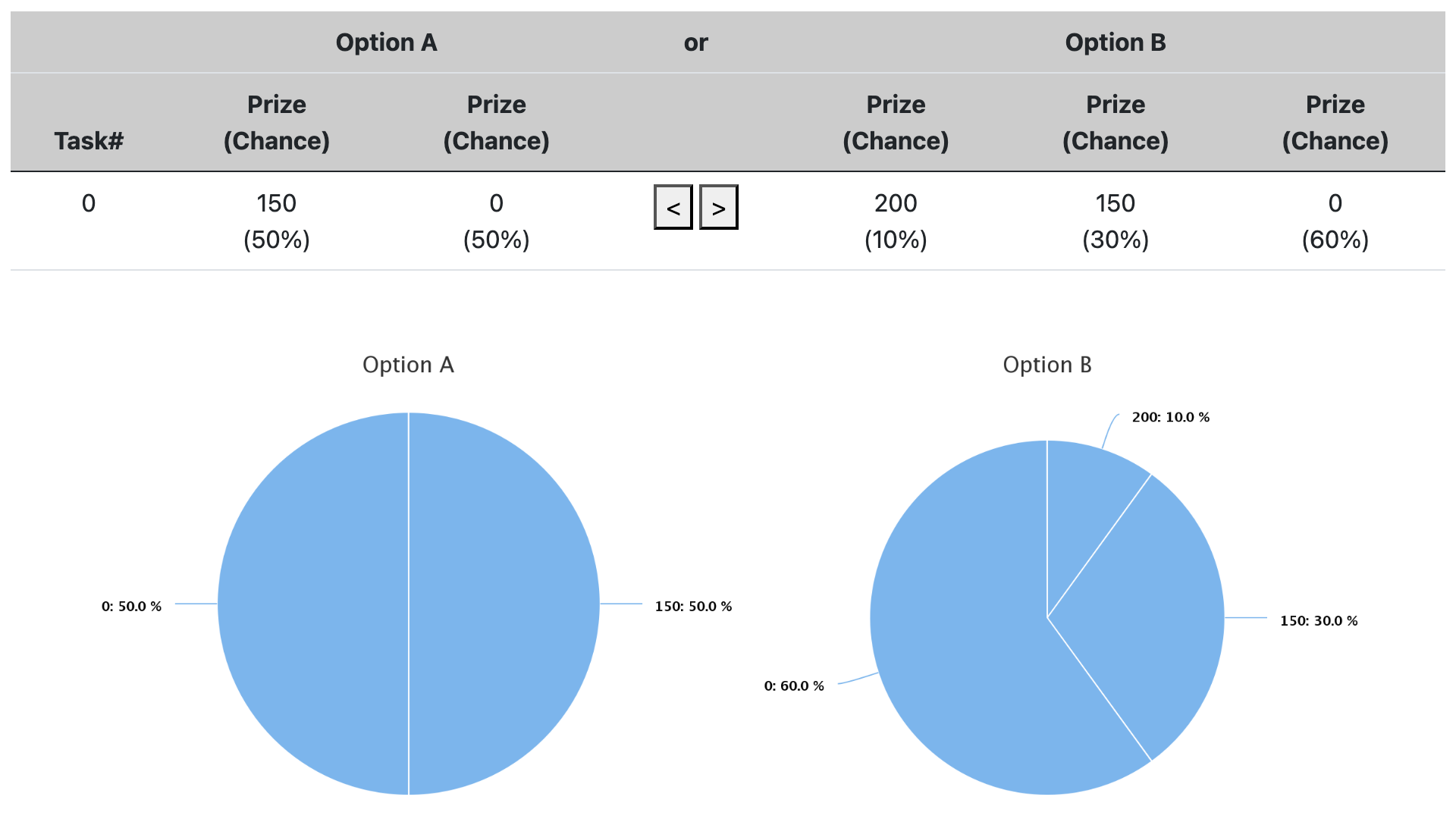}
\end{figure}

The columns and the image on the left represent a hypothetical Option A that gives you 150 points with 50\% chance and 0 points with 50\% chance.
The columns and the image on the right represent a hypothetical Option B that gives you 200 points with 10\% chance, 150 points with 30\% chance, and 0 points with 60\% chance.
Your task is to indicate which of the two options you prefer by clicking the arrow buttons located between the two lotteries.

The combination of options and the chances associated with them vary across tasks.
The amount of points you may earn in each Option varies from 0 to the maximum of 800. The chances associated with them vary from 0\% to 100\%.

\begin{figure}[ht]
    \centering
    \includegraphics[width=0.9\textwidth]{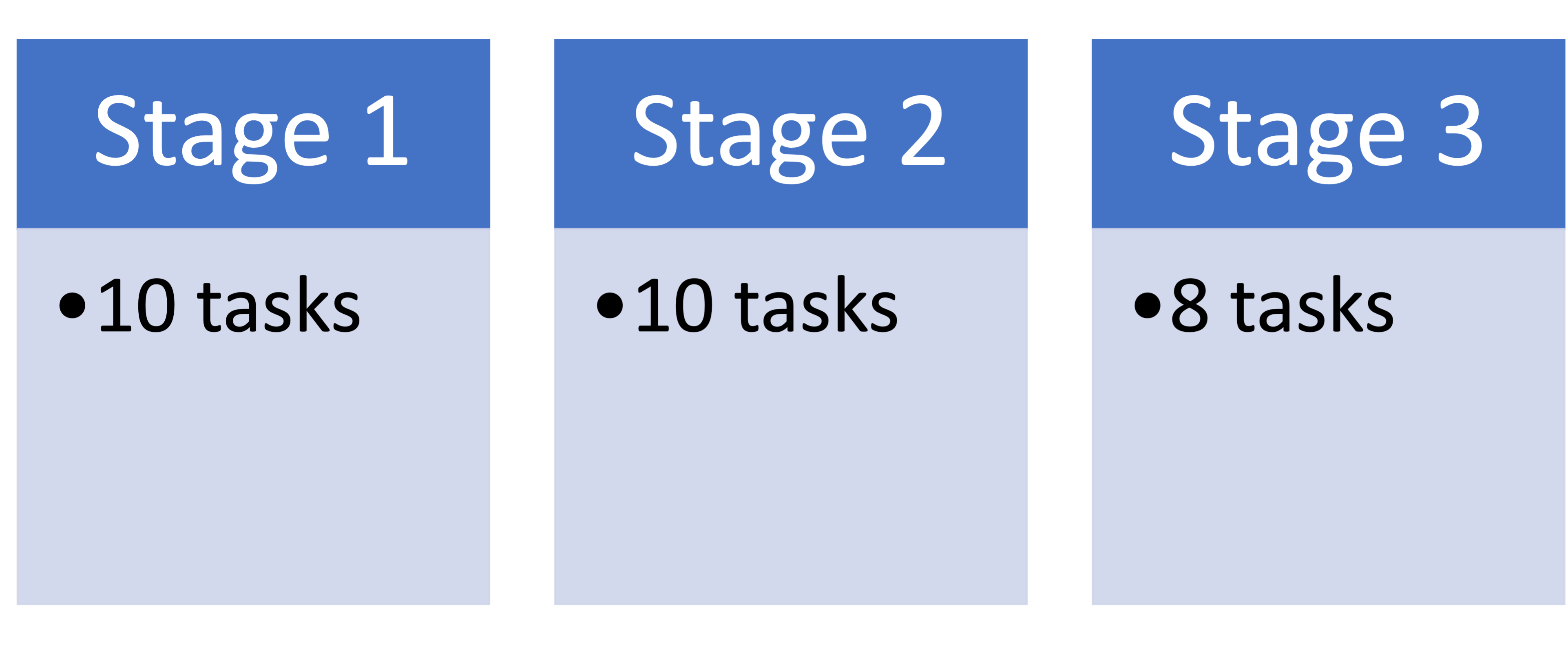}
\end{figure}

There are three stages in today’s experiment.
In Stage 1 and Stage 2, we present a table that contains a series of ten (10) tasks.
Each row represents a task of choosing between Option A and Option B.

In Stage 1, we will fix the chances and vary the amounts of points you can receive in each Option across the ten rows.
In Stage 2, we will fix the amounts of points you can receive and vary the chances associated with them across the ten rows.

In Stages 1 and 2, we will ask you to simply indicate a threshold line above which you would rather have Option A and below which you would rather have Option B, and vice versa. 
You can do so by clicking one of the arrows displayed in the center column of the table. 
Once you make your choice, a confirmation prompt will appear at the bottom of the screen to remind you of your selection. 
This will simplify the experiment and reduce the amount of repetitive procedures you need to complete. 
By doing so, you will only need to perform this procedure once instead of ten times for each row.

In Stage 3, we will present one task at a time. There will be eight (8) tasks in Stage 3, and we will vary both the amounts of points you can receive and the chances associated with them.

Combining all three Stages, there are total of 28 tasks in today’s experiment.
After completing all 28 tasks, the computer will randomly select one task for the payment.
Each task has an equal chance of being selected.

The final amount you receive will depend on your choice in the randomly chosen task.
The computer will simulate a lottery draw based on the option of your choice.

For example, if the option of your choice were Option A in the hypothetical example, you will receive 150 points with 50\% chance and 0 points with 50\% chance.

For your convenience, imagine this drawing process as the following.
Suppose there is a box that contains 50 red balls, and 50 yellow balls.
The computer shuffles the box and take one ball out at random.
If the ball is red, you receive 150 points, hence 50\% chance.
If the ball is yellow, you receive 0 points, hence 50\% chance.

If the option of your choice were Option B, on the other hand, you will receive 200 points with 10\% chance, 150 points with 30\% chance, and 0 points with 60\% chance.

For your convenience, imagine this drawing process as the following.
Suppose there is a box that contains 10 red balls, 30 yellow balls, and 60 green balls.
The computer shuffles the box and take one ball out at random.
If the ball is red, you receive 200 points, hence 10\% chance.
If the ball is yellow, you receive 150 points, hence 30\% chance.
If the ball is green, you receive 0 points, hence 60\% chance.

We will convert 100 points to \$1.00. That is, each point you receive is worth \$0.01.

This is the end of the instructions.

You may use the 'Previous' button to read the previous pages. Please make sure you understand the instructions well before starting the quiz.

When you are ready, please click the next button to start the quiz.

\end{document}